\documentclass[structabstract]{aa}
\usepackage{graphicx, xspace, lscape}
\usepackage{float}
\usepackage{longtable}
\usepackage{savesym}
\usepackage{amsmath}

\usepackage{txfonts}

\def\kms{$\mbox{km~s}^{-1}$\xspace}

\begin{document}
   \title{A VLT/FLAMES survey for massive binaries in Westerlund~1: VII. Cluster census\thanks{This work is 
     based on observations collected at the European Southern Observatory, 
     Paranal Observatory under programme IDs ESO 081.D-0324, 383.D-0633, 087.D-0440 and 091.D-0179}}
\author{J.~S.~Clark\inst{1} \and B.~W.~Ritchie\inst{1,2}  
      \and I.~Negueruela\inst{3}}
\institute{
$^1$School of physical sciences, The Open University, Walton Hall, Milton Keynes, MK7 6AA, United Kingdom \\
$^2$Lockheed Martin Integrated Systems, Building 1500, Langstone, Hampshire, PO9 1SA, UK \\
$^3$Departamento de F\'{\i}sica Aplicada, Facultad de
Ciencias, Universidad de Alicante, Carretera San Vicente  s/n,
E03690, San Vicente del Raspeig, Spain}

   \abstract
    {The formation, properties, and evolution of massive stars remain subject to considerable theoretical and observational uncertainty; impacting on fields as diverse as galactic feedback, the production of cosmic rays, and the nature of the progenitors of both electromagnetic and gravitational wave transients.}
   {The  young massive clusters many such stars reside within  provide a unique laboratory for addressing these issues, and in 
this work we provide a comprehensive stellar census of Westerlund 1, potentially the most massive Galactic example of such an aggregate, to underpin such efforts.}
   {We employ  optical spectroscopy of a large sample of early-type stars 
    to determine cluster membership for photometrically-identified candidates, characterise 
    their spectral type, and identify new candidate spectroscopic binaries.}
  {69 new members of Westerlund 1 are identified via $I-$band spectroscopy. Together with previous observations, they 
illustrate a smooth and continuous morphological sequence from late-O giant through to OB supergiant. Subsequently, the progression 
bifurcates, with one branch yielding 
mid- to late-B hypergiants and  cool super-/hypergiants, and the other massive blue stragglers, prior to a diverse population of H-depleted Wolf-Rayets.  
 We identify a substantial population of O-type stars with very broad Paschen series lines, a morphology 
    that is directly comparable to known binaries in the cluster. In a few cases additional
    low-resolution $R-$band spectroscopy is available, revealing
    double-lined He~I profiles and  confirming binarity for these objects; suggesting a correspondingly high
binary fraction amongst relatively unevolved cluster members.}
{Our current census remains incomplete, but indicates that Westerlund 1 contains at least 166
 stars with initial masses estimated to lie between $\sim25M_{\odot}$ to $\sim50M_{\odot}$, with more massive stars 
already lost to supernova. Our data is consistent with the  cluster being co-eval, although binary 
interaction is clearly required to yield the observed stellar population, which is characterised by a uniquely 
rich cohort of hypergiants ranging from  spectral type O to F,  with both mass-stripped 
primaries and rejuvenated secondaries/stellar mergers present. Future observations of Wd1 and similar stellar aggregates hold out the prospect of characterising both single- and  binary- evolutionary 
channels for massive stars and determining their relative contributions. This in turn will permit
the physical properties of such objects at the point of core-collapse to be predicted; of direct relevance 
for understanding the formation of relativistic remnants such as the  magnetars associated with Wd1 and other young massive clusters.}
  \titlerunning{Stellar census of Wd1}
  \maketitle
%

\section{Introduction}

Even before the detection of gravitational waves from coalescing neutron stars and black holes, the case for fully understanding the evolutionary cycle of massive stars was compelling. They, and their explosive endpoints contribute to the secular evolution of galaxies via radiative and  mechanical feedback and the deposition of dust and the products of nuclear burning into the wider
interstellar medium, facilitating subsequent generations of star formation. Upon core-collapse they yield a rich variety of energetic electromagnetic transients, with their relativistic remnants subsequently forming both X- and $\gamma$-ray binaries; phenomena that have also been associated with the production of cosmic rays. 

Rich young stellar clusters and associations potentially serve as powerful laboratories for investigating the properties and lifecycles of massive stars, as illustrated by  Massey et al. (\cite{massey00}, \cite{massey01}). 
Criticially, modern multiplexing spectroscopic instruments match the scale of such aggregates, meaning  
stellar populations can be efficiently categorised and, as a consequence,  evolutionary models observationally tested. Moreover, observations of multiple 
clusters at different ages means that we can reconstruct  the evolutionary pathways of stars sampling a wide range of initial 
masses. 

The efficacy of this methodology is evidenced by the investigation of the 30 Dor star formation region initiated by Evans et al. (\cite{evans}). However, as part of this multi-work campaign,  Schneider et al. (\cite{schneider18}) reveal a potential impediment to this approach; specifically 30 Dor is found to have a complex, extended formation history resulting in multiple spatially unsegregated stellar populations. Such a situation complicates the interpretation of evolutionary sequences, which are better constrained in simple, co-eval stellar populations. Moreover, at a distance of $\sim50$kpc essential observational diagnostics - such as long wavelength continuum observations to constrain mass loss rates - are difficult to apply.

Consequently, the identification of such stellar aggregates within the Galaxy would be of considerable value. 
As potentially the most massive open cluster within the Galaxy, Westerlund 1 (henceforth Wd1; Westerlund \cite{w61}, \cite{w87})
is a compelling target and, although heavily reddened, it remains accessible in the $R-$ and
$I-$bands. With an age of $\sim5$Myr, it  hosts a unique population of cool super-/hypergiants in addition to  significant numbers of OB supergiants, Wolf-Rayet (WR) stars and the magnetar CXO J164710.2-455216 (Clark et al. \cite{clark05}, Crowther et al. \cite{crowther06}, Muno et al. \cite{muno06a}, Negueruela et al. \cite{neg10}, Kudryavtseva et al. \cite{kud}).

Wd1 is spectacular in appearance across the electromagnetic spectrum, with point source (stellar) and diffuse emission observed at:

\begin{itemize}

\item X-ray energies due to emission from the magnetar, shocks in the winds of single stars, the wind collision zones of massive binaries, (unresolved) pre-main sequence  low-mass stars
and the hot intracluster medium  (Clark et al. \cite{clark08}, \cite{clark19c}, Muno et al. \cite{muno06b}).
Spatially extended,  highly  energetic $\gamma$-rays (GeV and TeV; Ohm et al. \cite{ohm} and Abramowski et al. \cite{abramowski}, respectively) have also been associated with Wd1 and as a consequence it has also been implicated in the production of cosmic rays. Both phenomena have been attributed to the 
(interactions between the) winds of massive single and binary stars, the cluster wind and supernovae  (e.g. Aharonian et al. \cite{aharonian}, Bednarek et al. \cite{bed}, Bykov et al. \cite{bykov}, Cesarsky \& Montmerle \cite{CM}).

\item IR emission is due to warm dust associated with interacting O-star and Wolf-Rayet binaries and  the ejection nebulae of 
cool super-/hypergiants, which is subsequently entrained within the intracluster medium  (Clark et al. \cite{clark98}, \cite{clark13}, Crowther et al. \cite{crowther06}, Dougherty et al. \cite{dougherty}).

\item mm- and radio-continuum emission is attributed to stellar winds from early-type and WR stars, circumstellar 
ejecta associated with the cool stellar cohort, wind collision zones in interacting binaries and the ionised intracluster 
medium (Dougherty et al. \cite{dougherty}, Fenech et al. \cite{fenech17}, \cite{fenech18}, Andrews et al. \cite{andrews}).

\end{itemize}

These phenomena are all ultimately driven by the underlying massive stellar population of the cluster in both life and death; hence if we are to understand them we are required to construct an accurate stellar census for Wd1. Since many are also manifestations of binary interaction, any such efforts should also serve to constrain the nature of this population, noting that multi-wavelength (X-ray, IR and radio) and multi-epoch photometric observations imply a high binary fraction (Clark et al. \cite{clark19c}, Crowther et al. \cite{crowther06}, Dougherty et al. \cite{dougherty} and Bonanos \cite{bonanos}, respectively). Such a rigorously determined population census is also 
essential in order to constrain the bulk properties of the cluster such as age, mass, initial mass function, degree of mass segregation, dynamical state  and mode of formation (e.g. Andersen et al. \cite{andersen}, Brandner et al. \cite{brandner}, Cottaar et al. \cite{cottaar}, Gennaro et al. \cite{gennaro11}, \cite{gennaro17}, Kudryavtseva et al. \cite{kud}, Lim et al. \cite{lim}, Negueruela et al. \cite{neg10}).   

Consequently between 2008-2013 we undertook a large multi-epoch $I-$band spectroscopic survey of massive stars within Wd1 with the Fibre Large Array Multi Element
Spectrograph (FLAMES; Pasquani et al. \cite{pasq02}) mounted on the Very Large Telescope (VLT). These observations were designed to (i) constrain the binary population of Wd1 via radial velocity (RV) monitoring, (ii) provide an expanded stellar census encompassing the  fainter cohort of massive evolved stars and (iii) supplement extant spectroscopic and photometric data in order to permit quantitative model atmosphere analysis of individual cluster members. To date these observations have allowed us to investigate the incidence of binarity amongst the supergiants (Ritchie et al. \cite{ritchie09a}, \cite{ritchie11}), 
provide tailored analyses of a number of interacting and post-interaction systems (Clark et al. \cite{clark11}, \cite{clark14b}, \cite{clark19a}, Ritchie et al. \cite{ritchie10}) and relate the X-ray properties of cluster members to their underlying stellar properties (Clark et al. \cite{clark19c}). In this work we present an enlarged census for Wd1, comprising 166 massive evolved stars, and discuss the implications of this population for both single- and binary-star evolutionary channels,  and the   nature of the cluster -  both in isolation and in comparison to other  young massive stellar aggregates.


\section{Observations \& data reduction} 
\label{sec:obs_data}

As described above the original, and primary, science goal of our programme was the identification
and characterisation of massive binaries within Wd1. In order to accomplish this 
multiple observations of four target fields were made in service mode
during 2008 and 2009 using the FLAMES-GIRAFFE multi-fibre spectrograph
with setup HR21 to cover the 8484-9001{\AA} range with
$R=\lambda/\Delta\lambda \sim 16200$. These comprised a \textit{bright} field, 
containing 22 spectroscopically- and photometrically-selected targets 
(detailed in Ritchie et al. \cite{ritchie09a}) and three \textit{faint}
fields containing 17 spectroscopically-confirmed cluster members and
63 photometrically-selected candidates. With three exceptions - Wd1-57a, Wd-71 and WR K -  the \textit{faint}
lists contain only stars with previously-known spectral types no later
than B0~Iab, or photometry consistent with lower-luminosity O9-B0
stars just evolving towards the supergiant phase (cf. Fig. 1).
Integration times were typically $2\times600$s for the \textit{bright}  and 
$3\times895$s for the three \textit{faint} configurations. 
Given the appearance  of some targets in more than one configuration  - motivated by the desire
to maximise fibre allocation - and the differing frequency with which each configuration was executed,
individual stars were ordinarily observed between five and eleven times.

Subsequently, additional observations were made in 2011 and 2013 utilising the same configuration. Target 
selection was optimised to include (i) follow up of spectroscopically and/or photometrically identified 
binary candidates and (ii) initial observations  of previously unclassified luminous  candidate cluster members 
and/or X-ray bright stars. Due to the observing runs remaining incomplete and the constraints imposed
by fibre allocation, the {\em final} number of observations  (and hence integration time) for individual 
stars varies considerably; a full observing log will be provided in Clark et al. (in prep.) where we 
analyse the full RV dataset.

The FLAMES-GIRAFFE pipeline and Common Pipeline Library  were used to
bias subtract, flat field and wavelength calibrate the
data, while subsequent processing made use of
IRAF\footnote{IRAF is distributed by the National Optical Astronomy
  Observatories, which are operated by the Association of Universities
  for Research in Astronomy, Inc., under cooperative agreement with
  the National Science Foundation.} for the extraction individual
spectra, rectification and heliocentric velocity correction; full
details of data reduction are given in Ritchie et al. (\cite{ritchie09a}).

Fortuitously, $R-$band spectra of a number of our targets were acquired on
the nights of 12 and 13 June 2004 using VLT/FORS2 in longslit and mask
exchange unit (MXU) modes, and are used to supplement the limited
wavelength coverage of these targets provided by our single VLT/FLAMES
setup. The G1200R grism was employed, yielding a nominal dispersion of 
0.38{\AA}/pixel over the spectral range 5750-7310{\AA}. A 0.3" slit was 
utilised for the longslit mode to obtain a resolution of 
$\sim7000$. For the MXU mode we employed 1" slits; these observations had a resolving power 
of $\sim2200$, with the exact wavelength coverage dependant on the position on the CCD. 
Data reduction for these objects is as described in Negueruela et al. 
(\cite{neg10}).

Finally additional  spectra in the outskirts of Wd1 were obtained with the fibre-fed dual-beam 
AAOmega spectrograph on the 3.9\,m Anglo-Australian Telescope (AAT) at the Australian Astronomical
Observatory on 2011, July 21. The Two Degree Field (``2dF") multi-object system
was utilised. Light from an optical fibre of diameter $2\farcs1$ on the sky is
fed into two arms via a dichroic beam-splitter with crossover at 5700{\AA}. Each
arm of the AAOmega system is equipped with a 2k$\times$4k E2V CCD detector and
an AAO2 CCD controller. Given the very high reddening to the targets, the blue
arm did not produce useful spectra. The red arm used the 1700D grating,
providing a resolution $R\sim11\,000$ in the region surrounding the Ca\,{\sc ii}
triplet. The spectra cover a $500${\AA} wide range centred on $8700${\AA},
but the projection of the spectrum from each fibre on the CCD depends on its
position on the plate, displacing the range limits up to $\sim$20{\AA}. The
exposure time was 1200 s. Data reduction was subsequently carried out
in the manner described in Gonz\'{a}lez-Fern\'{a}ndez et al. (\cite{CG}).

\begin{figure}
\begin{center}
\resizebox{\hsize}{!}{\includegraphics{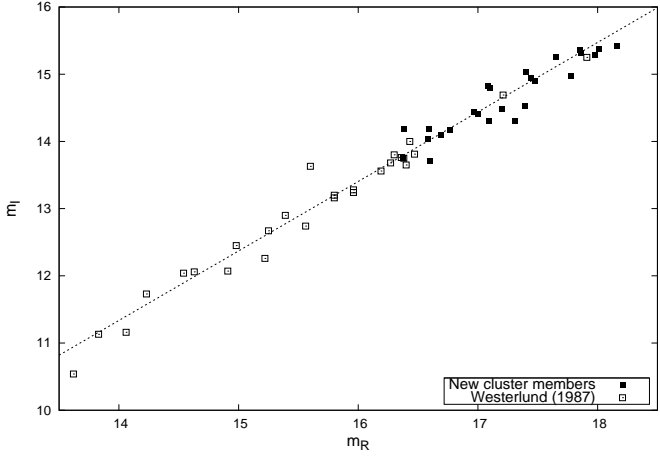}}
\caption{Comparison of $R-$ and $I-$band photometric magnitudes from Clark et al. (\cite{clark05}) for
  cluster members listed by Westerlund (\cite{w87}; open squares) and new
  confirmed cluster members listed in Table~2
  (solid squares). }
\label{fig:phot}
\end{center}
\end{figure}

\section{New cluster members}\label{sec:results}

\begin{figure*}
\begin{center}
\resizebox{\hsize}{!}{\includegraphics{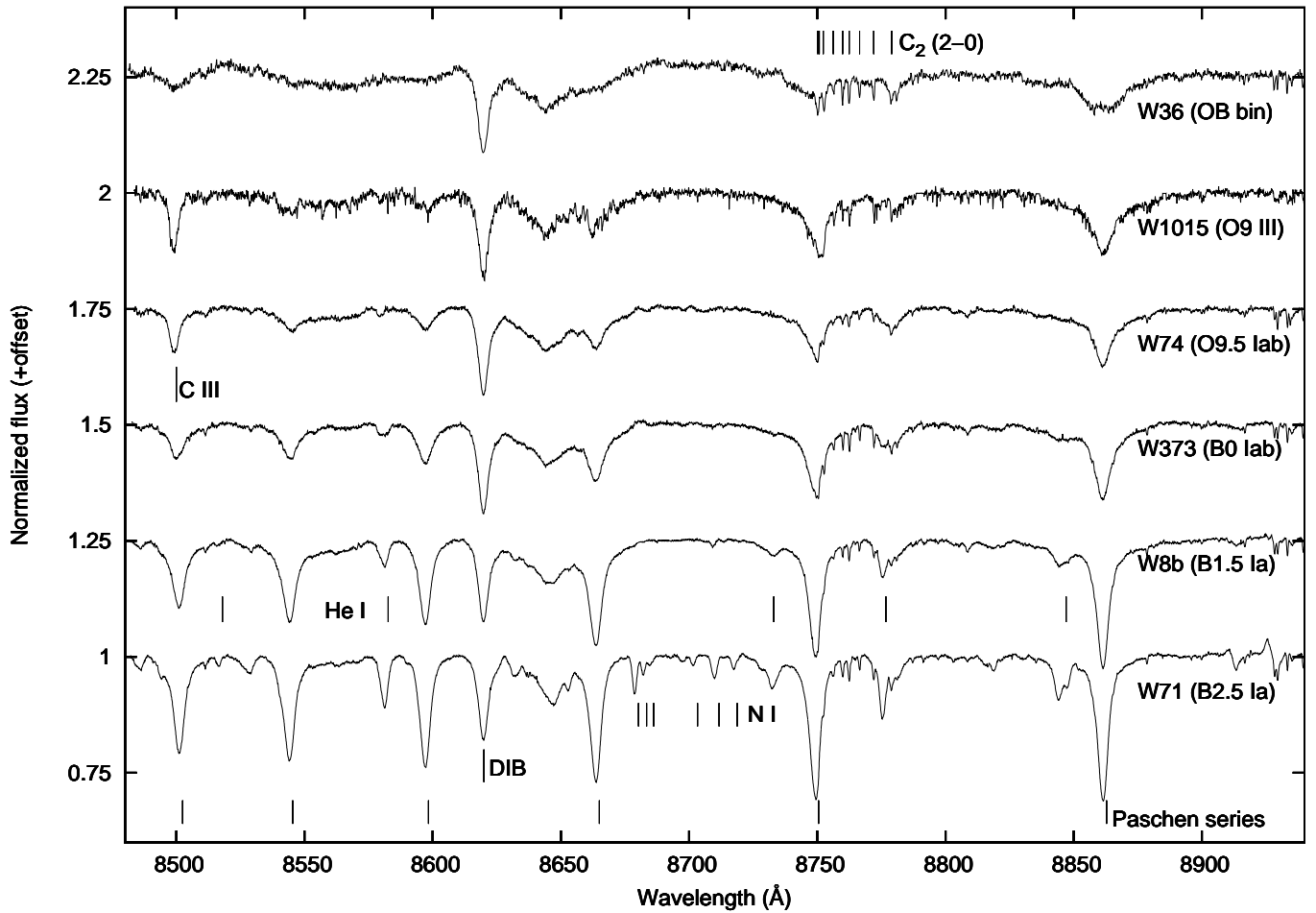}}
\caption{Montage of $I$-band spectra of classification standards for apparently single stars of spectral type O9-B2.5. The spectrum of the 3.18-day eclipsing binary Wd1-36 is included to indicate the appearance of SB2 systems of comparable spectral type within 
Wd1; note the shallow, flat-bottomed Paschen series lines (see Sect. 3.5).}
\label{fig:standards}
\end{center}
\end{figure*}

\begin{figure*}
\begin{center}
\includegraphics[width=13cm,angle=-90]{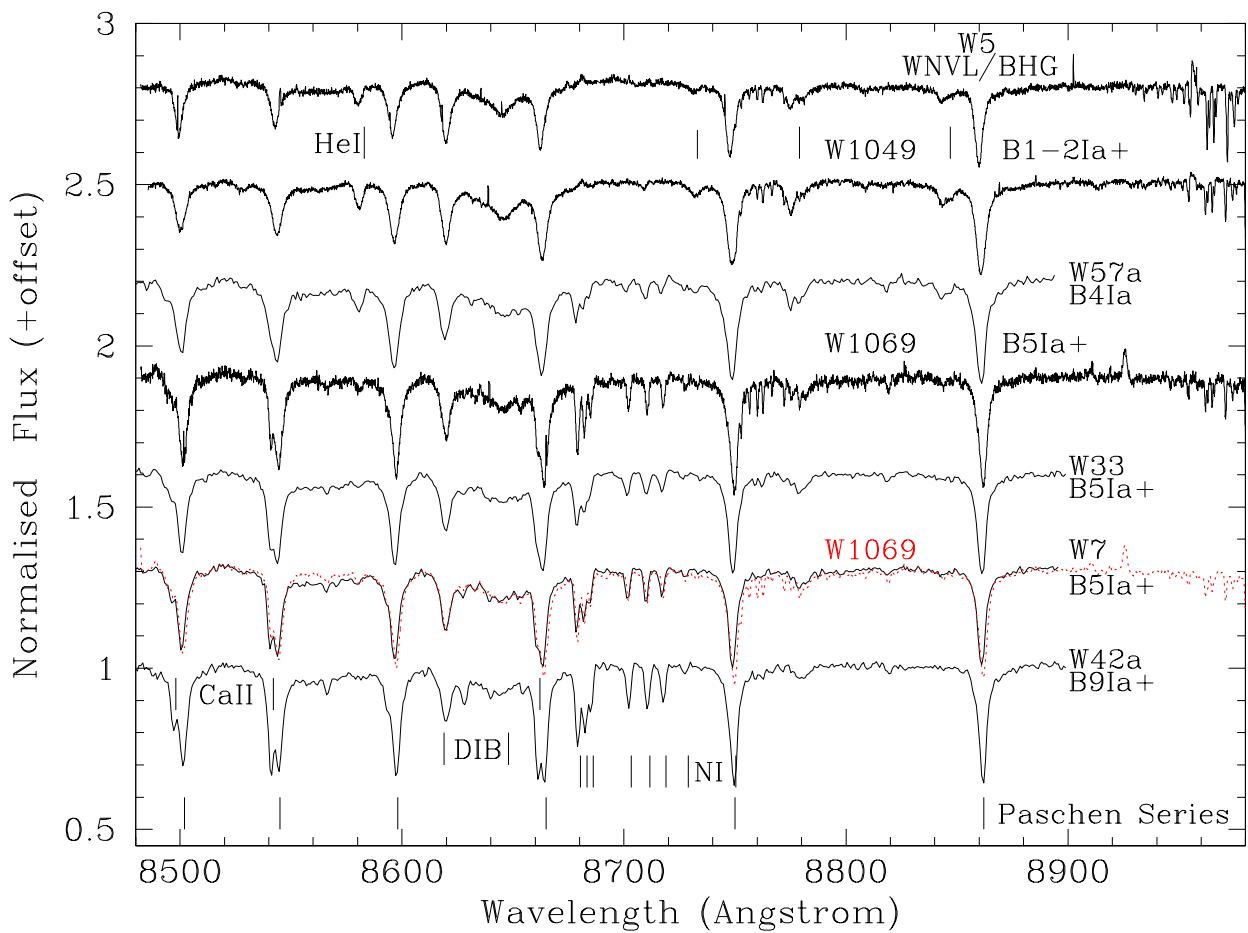}
\caption{New $I$-band spectra of W1049 and W1069 plotted against a  montage of cluster  B0-B9 hypergiants in order to 
permit classification (with the highly luminous B4 Ia star Wd1-57a included for completeness). The spectrum of W1069 has been degraded in resolution
(red dashed line) to enable direct comparison to the spectra of B5-9 hypergiants.}
\label{fig:standards}
\end{center}
\end{figure*} 

\begin{figure*}
\begin{center}
\includegraphics[width=13cm,angle=-90]{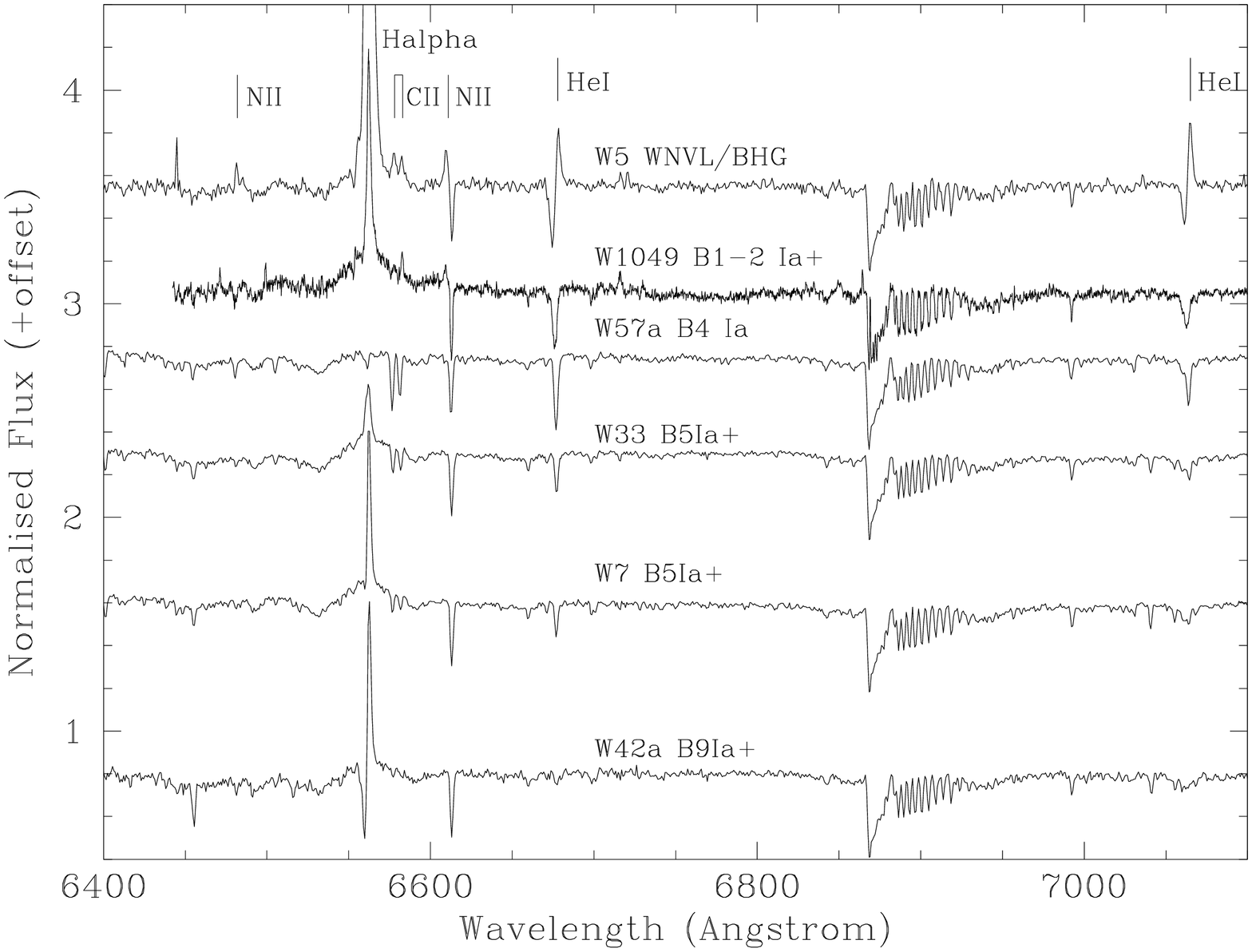}
\caption{$R$-band spectrum of W1049 plotted with comparable spectra of highly luminous cluster super-/hypergiants of spectral type B0-B9.}
\label{fig:standards}
\end{center}
\end{figure*} 

We are able to identify a total of 69 new cluster members via our spectroscopy. 
Foreshadowing Sect. 3.1-3.5, the resultant stellar census for Wd1 
is summarised in Table 1, while the co-ordinates, photometry and spectral
classifications for individual members are presented in Table 2. Once ordered by increasing right 
ascension, we apply a
$W1xxx$ designation for these new sources\footnote{We suggest that this naming convention supercede that employed 
for the subset  of these stars described in previous works (e.g. Ritchie et al. 
\cite{ritchie09a}, Fenech et al. \cite{fenech18}). To enable cross-correlation, 
where appropriate, we give the old nomenclature in Table 2.}.

 In addition to these stars, eight further objects were observed.
Seven of these eight targets were found to be late-type stars\footnote{
Three of these stars - F1 (see Ritchie et al. \cite{ritchie09a}), F2 and F4 - show a TiO 8860{\AA}
 bandhead (Ramsey \cite{ramsey}) with a strength implying they are
early-M giants. The remaining objects show spectra typical of late-G
or early-K stars, with strong Ca\,{\sc ii} and Mg\,{\sc i} 8807{\AA} and 8837{\AA} lines
and Paschen series absorption that is either very weak or absent.}, while the
final object is a B-type star of luminosity class V. 
The RVs measured for the majority of these indicate that they
are foreground objects, but two objects, F7 and
F8, have RVs consistent with membership of Wd1 (Clark et al. \cite{clark14b}, in prep.). However, the
pronounced DIB at $\sim$8620{\AA} that is characteristic of {\em bona fide} 
cluster members (cf. Sect. 3.1)  appears weak or absent for both stars, which are also spatial outliers
to the east and north of the cluster respectively. Given their apparent classification
as mid-K giants we conclude that these are also interlopers.

\subsection{Spectroscopic classification}
\label{sec:classification}

\begin{table}
\caption{Stellar demographics of Wd1}
\begin{center}
\begin{tabular}{lcc}
\hline
\hline
Spectral    & Cluster     &  New to  \\
Type        & population  &  this work  \\
\hline
O9-9.5 III             &   27  & 25 \\
O9-9.5 II,II-III       &   11  & 11 \\
O9-9.5 Iab,Ib          &   31  & 11 \\
B0-0.5 Ia,Iab,Ib        &   23  & 7 \\
B1-1.5 Ia,Iab   & 10     & 2    \\
B2-4 Ia         & 7 & 0 \\[2.5mm]
O4-8 Ia$^+$     & 2 & 0 \\
B0-2Ia$^+$/WNVL & 4 & 1 \\
B5-9 Ia$^+$     & 4 & 1 \\[2.5mm]
LBV             & 1 & 0 \\
YHG+RSG         & 10 & 0 \\[2.5mm]
sgB[e]          & 1 & 0 \\
OB SB2          & 12 & 10 \\[2.5mm]
OeBe            &  1 & 1 \\[2.5mm]
WN5-8           & 14 & 0 \\
WC8-9           & 8 & 0 \\
\hline
Total           & 166 & 69 \\
\hline
\end{tabular}
\end{center}
{Sub-panels in the table comprise, from the top: giants and supergiants, hypergiants, transitional stars, binaries
of uncertain spectral type and/or luminosity class (e.g. Wd1-36, W1001, W1003, and the supergiant B[e] (sgB[e]) star Wd1-9), OeBe stars 
and Wolf-Rayets. Transitional stars comprise luminous blue variables (LBVs), yellow hypergiants
(YHGs), and red supergiants (RSGs). We include the O9-9.5 I-III stars W1033 and -1040 in the O9-9.5 Iab,Ib 
category (Sect. 3.4); thus explaining the slight discrepancy between this table and the version in Clark et al. (\cite{clark19c}).}
\end{table}

Co-added spectra of (apparently) non-binary objects are used as
standards for classification of the newly-identified stars in Wd1,
with an appropriate $I-$band stellar sequence shown in Figs. 2 and 3. 
Following guidelines to classifying
O9--B9 stars, taken from Clark et al. (\cite{clark05})  and
Negueruela et al. (\cite{neg10}), the following diagnostics are used:

\begin{itemize}

\item The comparative weakness of the higher Paschen lines and the strength of
  C\,{\sc iii} 8500{\AA} leads to a ratio
  \mbox{(Pa-16+C\,{\sc iii} 8500{\AA})/Pa-15$\gg$1} for O-type stars. At O9
  the Pa-15 line is hard to distinguish from the continuum.

\item At B0, Pa-15 and Pa-16 lines are of approximately equal strength
  due to the increasing contribution of C\,{\sc iii} 8500{\AA} to the
  Pa-16 line. He\,{\sc i} 8583{\AA} is still clear, while
  He\,{\sc i} 8847{\AA} is seen as an inflection on the blue wing of
  Pa-11.

\item B0.5-2 spectral types show a rapid increase in the strength of
  both the Paschen series and the He\,{\sc i} lines (Negueruela et al. \cite{neg10}), 
  with all objects later than B0 showing \mbox{Pa-15/Pa-16$>1$}. A wealth of N\,{\sc i} 
lines between $\sim8680-8730${\AA} begin to appear at $\sim$B2 and are clear by B2.5\footnote{N\,{\sc i}
8680.4{\AA}, 8683.4{\AA}, 8686.1{\AA}, 8703.2{\AA}, 8711.7{\AA}, 8718.8{\AA} and 8728.9{\AA}.}

\item For spectral types B2.5 and later both the Paschen series and N\,{\sc i} photospheric
lines continue to increase in strength, while the He\,{\sc i} lines drop out by $\sim$B5. At the same point, comparison of the 
strength of the emerging Ca\,{\sc ii} 8498.0{\AA}, 8542.1{\AA} and 8662.1{\AA} absorption features to their adjacent Paschen series lines allows spectral types to $\sim$B9 to be distinguished.

\end{itemize}

With our limited spectral coverage and paucity of lines suitable for
classification, distinguishing between O9 and O9.5 is challenging.  At
high $S$/$N$ He\,{\sc i} 8583{\AA} remains visible at O9.5, but generally
cannot be distinguished at the lower $S$/$N$ of many faint targets,
and C\,{\sc iii} 8500{\AA} line strength represents the best available
classifier, while the width and profile of the wings of Pa-11 serve as
a guide to the luminosity class. However, as discussed in
Negueruela et al. (\cite{neg10}), these features will also be affected by rotation and
abundance anomalies, and classifications given here are therefore
uncertain to $\sim$half a spectral type.

A further potential difficulty in classifying O-type stars is introduced
by binarity. The brevity of  (post-MS) evolutionary phases means that a system
containing a luminous supergiant primary will appear single-lined
unless the mass ratio is near unity -- such that both objects are in
a comparable state -- and will therefore display a spectrum that is
very similar to that of a single star.  
 However systems that consist of an $\sim$O9--9.5~III primary and massive OB
secondary are more likely to display composite spectra.
Consideration of synthetic spectra suggests that the
relatively broad Paschen series lines in a short-period O9~III+O9~III
system would remain heavily blended even at quadrature. 
We might therefore expect such systems to  display a spectrum with broad,  flat-bottomed
Paschen series lines and anomalously broad, weak C\,{\sc iii} 8500{\AA} that would prove difficult to classify.
An exampler of this phenomenon, albeit of greater luminosity,  is the  3.18-day eclipsing binary Wd1-36 
(Bonanos \cite{bonanos}); as a consequence of the strong line-blending we are only able to prove a 
generic OB Ia + OB Ia classification for this  system. This also serves to illustrate the significant  
morphological differences between the spectra of such systems and those of stars following the empirical 
O9 III - B2.5 Ia evolutionary sequence illustrated by Fig. 2; we discuss this and similar systems  in greater detail in Sect. 3.5.

A number of interstellar features are also visible in our spectra. The
strong, well-defined $\sim$8620{\AA} DIB has been discussed in Ritchie et al. (\cite{ritchie09a}),
and serves as a check for zero-point RV errors in our data; in 18
observations of the luminous B2.5~Ia supergiant Wd1-71, we find
$\sigma_\text{RV(8620)}=0.76$\kms and variations in FWHM of less than
2\%. A broad DIB at $\sim8648${\AA} is also apparent (see also Negueruela et al.
\cite{neg10} and references therein) and a DIB at 8529{\AA} is also
seen in high-$S$/$N$ spectra. A number of weaker interstellar features
are also present, with the C$_2$ Phillips (2--0) band system
(e.g. van Dishoeck \& de Zeeuw \cite{vddz84}) prominent on the red wing of the Pa-12
line. The $R$ lines largely overlap the core of Pa-12 and cannot be
measured with accuracy, but in very high $S$/$N$ co-added spectra of
Wd1-71 (total integration time $\sim$9~hours) the $Q$ lines are
very strong and may be distinguished to Q(18), while a strong R(0)
line, moderate P(2) and P(8) lines and weak R(10), R(12) may also be
distinguished, with P(4) and P(6) blended with the wings of
He\,{\sc i} 8777{\AA}. The mean radial velocity of the eleven lines that could
be measured with accuracy, weighted by equivalent width, is
$-44.7\pm1.9$\kms, very close to typical systemic velocities for the
stellar population of Wd1, suggesting that the absorbing material is
local to the cluster. The centre of the strong $\sim$8620{\AA} DIB lies
at 8619.8{\AA}; if this is assumed to originate in the same material as
the Phillips (2--0) system then the intrinsic DIB wavelength is in
good agreement with the value of 8621.2{\AA} reported by Jenniskens \& Desert (\cite{jd94}).

\subsection{Blue hypergiants}

 We turn first to W1049 and W1069; two new reddened hypergiant candidates identified via our AAT observations (Figs. 3 \& 4). W1069 has only been observed in the  $I-$band, but it appears to be an excellent match to the B5 Ia$^+$ cluster members Wd1-7 and -33; the strength of the Ca\,{\sc ii} lines precluding an earlier ($\leq$B4; Wd1-57a) or later ($\sim$B9; Wd1-42a) spectral type. By contrast the $I-$band spectrum of W1049 appears earlier, matching that of both the apparently single B1.5 Ia supergiant Wd1-8b (Fig. 2) 
and Wd1-5 (WN10-11h/B0.5 Ia$^+$; Crowther et al. \cite{crowther06}, Negueruela et al. \cite{neg10}). 

Unlike W1069 an $R-$band spectrum is available and, as with Wd1-5, W1049 appears to differ from the normal supergiants
 within Wd1 (Fig. 4). Specifically both the strength and profile of the H$\alpha$  line is atypical for such stars, while weak emission in the C\,{\sc ii} 6582{\AA} and N\,{\sc ii} 6611{\AA} lines is also anomalous (Negueruela et al. \cite{neg10}). Instead these features characterise the B0-1 Ia$^+$/WN10-11h stars Wd1-5 and -13, while the H$\alpha$ profile - comprising a strong, 
narrow central peak and broad wings - is also reminiscent of these stars and, indeed, the B5 Ia$^+$ stars Wd1-7 and -33 (Fig. 4 and 
Ritchie  et al. \cite{ritchie10}).
A similar H$\alpha$ profile is observed for Cyg OB2 \#12 (B3-4 Ia$^+$) and $\zeta^1$ Sco (B1.5 Ia$^+$); albeit with 
the C\,{\sc ii} lines in absorption in both stars and a prominent P Cygni component characterising the latter star (such that it
resembles the much cooler Wd1-42a; Clark et al. \cite{clark12}). 

The corresponence between W1049, Wd1-5, and -13 is not exact, with  P Cygni profiles of the 
He\,{\sc i} 6678{\AA} and 7065{\AA} lines in the latter stars absent in the former. Nevertheless their overal similarity  
provides a compelling case for a hypergiant classification for W1049. This is particularly interesting since both Wd1-5 and -13 appear to result from binary driven mass-stripping (Ritchie  et al. \cite{ritchie10}, Clark et al. \cite{clark14b}); to date no search for binarity has been undertaken for W1049.

\subsection{Lower luminosity B stars}
\label{sec:bsgs}

Nine of the newly classified cluster members present spectra consistent with early-B supergiants.
 Only two stars  appear to have spectral types later than B0; W1039  and W1048 
resemble  Wd1-78 (B1~Ia; Neguerueula et al. \cite{neg10}) and Wd1-8b (B1.5 Ia; Fig. 2) respectively, and hence
we adopt such  classifications for them.
The other seven objects -  W1005, -1009, -1053, -1055, -1065, -1067, and -1068 - all appear to be 
B0 supergiants of luminosity class Iab or Ib. Of these W1068  shows discrepant Ca\,{\sc ii} 
and Mg\,{\sc i} lines that are most probably due to blending with a foreground late-type star.
W1055 is of note since it presents anomalously strong
C\,{\sc iii} 8500{\AA} absorption and a broadening of the Paschen series lines at some
epochs that suggests the presence of an O-type companion (Clark et al. in prep.), 
W1065 was likewise  identified as a short-period RV-variable binary in Ritchie et al. (\cite{ritchie11}); 
strongly infilled H$\alpha$ and C\,{\sc ii} 7231{\AA} and 7236{\AA} emission confirming a luminous supergiant
classification (Negueruela et al. \cite{neg10}). As such these add to other known early B supergiant binaries
such as Wd1-6a and -52 (Bonanos \cite{bonanos}), Wd1-10 (Clark et al. \cite{clark19c}), and Wd1-43a (Ritchie et al. 
\cite{ritchie11}).

Critically, we find no B-type stars from lower luminosity classes - such as  B2~Ib-II or B0.5--1~II-III stars that
would correspond to an older population of $M_{\rm init}\sim15-20M_{\odot}$   - which would present 
as anomalously photometrically faint objects with broad but sharp troughed Paschen series absorption profiles.

However, a tenth early-type star, W1004, 
displays Pa-11\dots 16 lines in emission with double peaks  characteristic of a classical OeBe
star. Unfortunately, contamination of the sparse diagnostic features by emission from the 
circumstellar disc prevents accurate classification of the star and hence a precise 
evaluation of its evolutionary state. Irrespective of this shortcoming,  
despite a location 2\arcmin40\arcsec to the south-west  
($3.7(d/5\text{kpc})$~pc in projection)  the 
strong $\sim$8620{\AA} and $\sim$8648{\AA} DIBs in the spectrum of W1004 have profiles that are identical 
to OB stars in Wd1. Wisniewski \& Bjorkman (\cite{wis}) suggest an onset of the Be phenomenon
around $\sim5$Myr while, at solar metalicities,  a handful of Oe stars have been identified  (Negueruela et al. 
\cite{neg04}, Vink et al. \cite{vink}). As a consequence we consider that W1004 is potentially a   
member of Wd1 and one that could derive from the underlying  co-eval cluster population defined 
by the stars reported here.

\subsection{O stars}
\label{sec:ostars}

\begin{figure}
\begin{center}
\resizebox{\hsize}{!}{\includegraphics{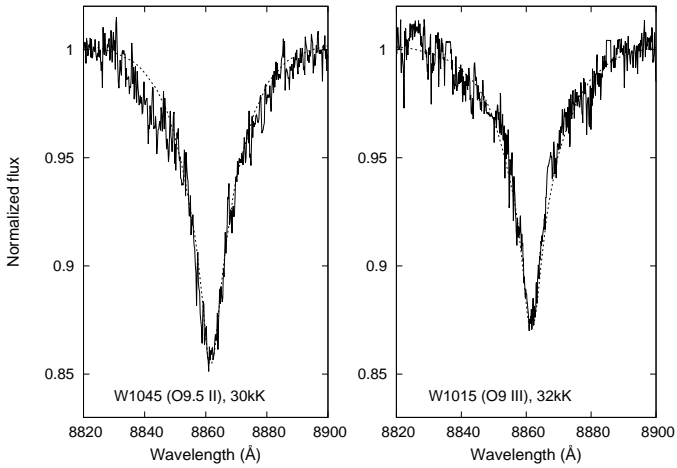}}
\caption{Comparison of the Pa11 line from W1045  (O9.5~II,
  left panel) and W1015 (O9~III, right panel) to synthetic model
  atmosphere spectra for 30kK and 32kK respectively. Note the
  influence of He\,{\sc i} 8839{\AA} on the blue wing of W1045, which is not
  well reproduced by the non-tailored fit used. }
\label{fig:synth}
\end{center}
\end{figure}

Finally we turn to the  majority of the new cluster members which, pre-empting the following discussion,
 we find to be  O-type stars of luminosity class III--Ib. Given the limitations of the                   
spectroscopic diagnostics present in the $I-$band we need to make explicit use of extant
photometry and, where available, $R-$band spectroscopy in order to obtain the most 
precise classifications for this cohort. Despite the possibility of a systematic offset in the classifications
we arrive at in comparison to those obtained from more traditional methodologies (e.g. employing blue-end spectral
diagnostics) we are confident our technique robustly  classifies such stars relative 
to other cluster members. Indeed it reinforces a central empirical finding of our survey; that there is a smooth
and continuous progression in spectral morphologies from the earliest O giants through to the mid-B supergiants and indeed the 
mid- to late-B hypergiants (cf. Figs. 2 and 3).

Nevertheless, while further spectroscopic and photometric observations of
the newly-identified cluster members are required to fully categorize
this population, in general the new cluster members fall into
four  distinct groups:
\begin{itemize}
\item Stars with spectra and photometry comparable  to Wd1-1 and -74 (O9.5 Iab;
Negueruela et al. \cite{neg10}). As a result these are given an identical classification.

\item Stars with spectra similar to Wd1-74, but with slightly broader Paschen series lines and
  (when photometry is available) $I-$band magnitudes roughly a magnitude
  fainter than the B0 supergiants listed in
  Table~2. These objects are classified as O9.5~II.

\item Stars with broader, weaker Paschen series lines than the O9.5~II
  stars, often with strong C\,{\sc iii}. In the few cases where photometry is
  available, these objects are 2--3 magnitudes fainter than O9.5--B0
  supergiants. These objects are classified as O9-9.5~III. 
Comparison with synthetic spectra generated with the non-LTE model
atmosphere code CMFGEN (Hillier \& Miller \cite{hillier98b}) suggests that such a 
classification is generally reasonable, with
\textit{non-tailored} fits at 30kK (O9.5) and 32kK (O9) shown in
Fig. 5.

\item Stars with very broad, weak Paschen series lines that are
  generally flat-bottomed. C\,{\sc iii} is also weak and broad. $R-$band
  spectra, when available, show shallow and possibly double-lined He\,{\sc i}
  and infilled H$\alpha$, while the objects may be unexpectedly
  luminous for the implied early spectral type. These objects have
  very similar I-band morphologies to known binaries in Wd1 and 
  are described in more detail in Sect. 3.5.
\end{itemize}

\subsubsection{O9-9.5 supergiants}

Seven stars\footnote{W1018, -1024, -1027, -1030, -1034, -1036, and -1064.} appear to be {\em bona
fide} late-O supergiants on the basis of their $I-$band spectra; we therefore adopt classifications 
of O9.5Iab for these objects (Table 2). W1036  has a $R-$band spectrum (Fig. 6) which,
 demonstrating strong, narrow H$\alpha$ and He\,{\sc i} photospheric absorption, is entirely 
consistent with this conclusion. 

However, a number of cluster members do not fit neatly into
this  scheme.  W1057  would also be classified as O9.5 Iab  on the basis 
of its $I-$band spectrum. However it is one of the more luminous O-type stars in our sample, with 
an $R-$band spectrum showing infilled H$\alpha$ and weak wind lines of C\,{\sc ii} (Fig. 6); the combination
of these properties suggest that it could instead be a rapidly-rotating B0~Iab
supergiant. As such we adopt a classification of O9.5--B0~Iab for W1057, noting that other O9.5Iab stars, such as
W1024 and -1034, are similarly luminous (although no $R-$band data are available for them).
The spectra of   W1033 and W1040  (=\object{C07-X5} and \object{C07-X3} respectively; Clark et al.
\cite{clark08}, \cite{clark19c}) are hard to interpret, with Paschen series lines that suggest a
classification of O9~III, but weak C\,{\sc iii} suggesting a later O9.5 (or
possibly B0) spectral type and photometric magnitudes that are
consistent with supergiants in Wd1. Low-resolution $R-$band spectroscopy
of W1040 , showing infilled H$\alpha$ (Fig. 6), again supports a
supergiant identification. We therefore assign  an intermediate classification of
O9-9.5 I-III for  both stars. Finally W1041 is classified as O9.5Iab but presents anomalously broad Paschen lines; 
we discuss this further in Sect. 3.5.

\begin{figure*}
\begin{center}
\resizebox{\hsize}{!}{\includegraphics{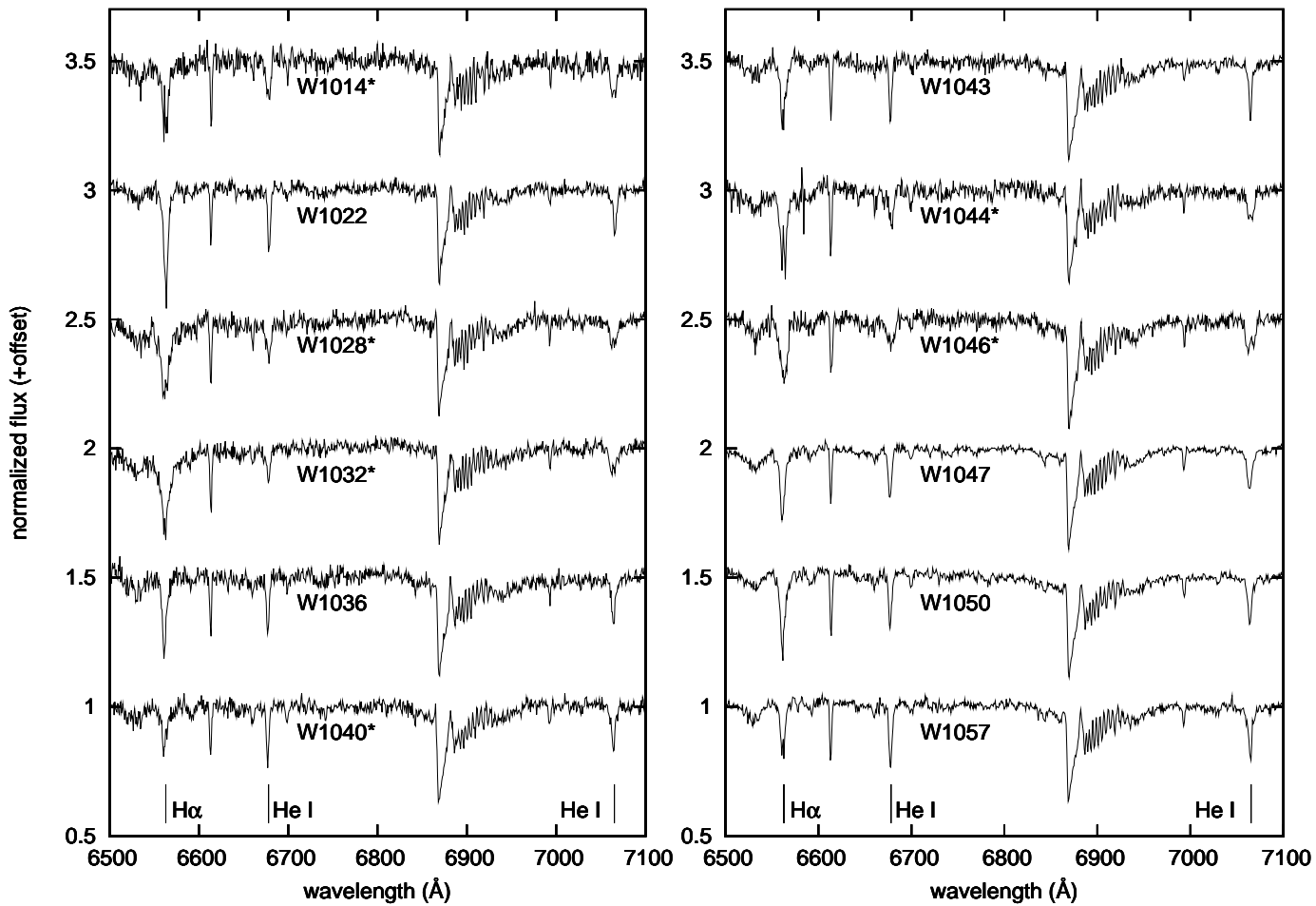}}
\caption{Montage of select FORS2/MXU $R$-band spectra of newly-identified
  cluster members. Objects marked with a `*' show very broad
  Paschen series lines indicative of a SB2.}
\label{fig:fors}
\end{center}
\end{figure*}

\subsubsection{O9.5~II stars}

Nine cluster members\footnote{W1008, -1022, -1037, -1042, -1045
-1047, -1050, -1056, and -1060} are classified as O9.5~II due to their
 spectroscopic similarity to Wd1-74 (O9.5~Iab), with the important diagnostic exceptions of
broader wings to the Paschen series lines  - indicative of a higher surface gravity - 
and $I-$band magnitudes that are somewhat lower than the O9.5~Iab supergiants. 
Low-resolution FORS2/MXU $R-$band spectra  are available for W1022, -1047  and -1050  (Fig. 6), all of 
which are consistent with such a classification,  demonstrating
 narrow single-lined photospheric He\,{\sc i} lines with H$\alpha$ in absorption.

W1043  is photometrically fainter than  this sample, with  Paschen series lines that 
also appear slightly broader; as such we assign a
provisional classification of O9.5~II--III. An $R-$band
spectrum is available (Fig. 6); He\,{\sc i} absorption lines are comparable to other O9.5~II stars, 
although H$\alpha$ appears somewhat broader and infilled. W1002  also has a luminosity similar 
to the O9.5~II systems discussed above but displays very broad, dilute Paschen
series lines that are morphologically similar to known binaries
in Wd1; given this  we suspect this system may contain an O9--9.5~II
primary and an O-type secondary, which we discuss further in Sect. 3.5.

Finally, Bonanos (\cite{bonanos}) reported that both W1002 and -1022 display
 rapid ($P$$<$0.2~day), periodic photometric variations and, in the absence
of spectroscopy, suggested that both are  $\delta$~Scuti variables. However
our  spectral classifictions are clearly inconsistent with such a suggestion 
and are also earlier than expected for $\beta$~Cephei-type variables
(cf. Pamyatnykh \cite{pam}).

\subsubsection{O9-9.5~III stars}

 After accounting for the above stars, 35 cluster members remain to be classified. Of these 12 stars\footnote{W1007, -1015, -1023, -1026, -1031, -138, -1051, -1052, -1058, -1059, -1063, and -1066.} fulfill the preceding observational criteria for classification as O9-9.5 III stars (cf. W1015; Figs. 2 \& 5). Unfortunately no $R-$band spectra are available for any of these objects.  
A further 13 stars\footnote{W1006, -1012, -1014, -1016, -1017, -1019, -1021, -1028, -1029, -1032, -1035, -1044, and -1061.} have spectra and, where available, photometry consistent with an O9-9.5 III classification - which we adopt - albeit with notably broader photospheric Paschen series lines. We discuss these stars as well  as (i) the handful of more luminous stars previous highlighted as 
showing the same phenomenon\footnote{W1002, -1040, -1041 and -1055} and  (ii) ten further, but more extreme 
examples\footnote{W1001, -1003, -1010, -1011, -1013, -1020, -1025, -1046, -1054, and -1062.}
- which, as a consequence defy precise determination of spectral type and/or luminosity class (cf. Tables 1 \&  2) - immediately below.

\subsection{Broad-lined O-type stars}
\label{sec:double}

\begin{figure}
\begin{center}
\resizebox{\hsize}{!}{\includegraphics{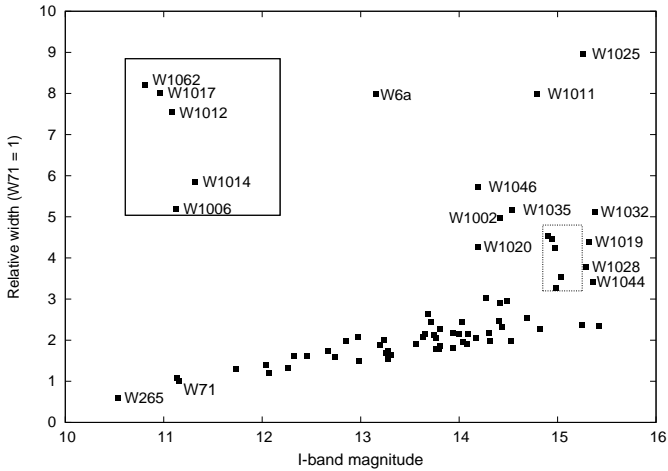}}
\caption{Relative Pa-11 line width as a function of $I-$band luminosity
  for targets with photometry from Clark et al. (\cite{clark05}). For clarity, the box in the top left of this figure 
presents an expanded view of the points within the small dotted region.}
\label{fig:widths}
\end{center}
\end{figure}

\begin{figure*}
\begin{center}
\resizebox{\hsize}{!}{\includegraphics{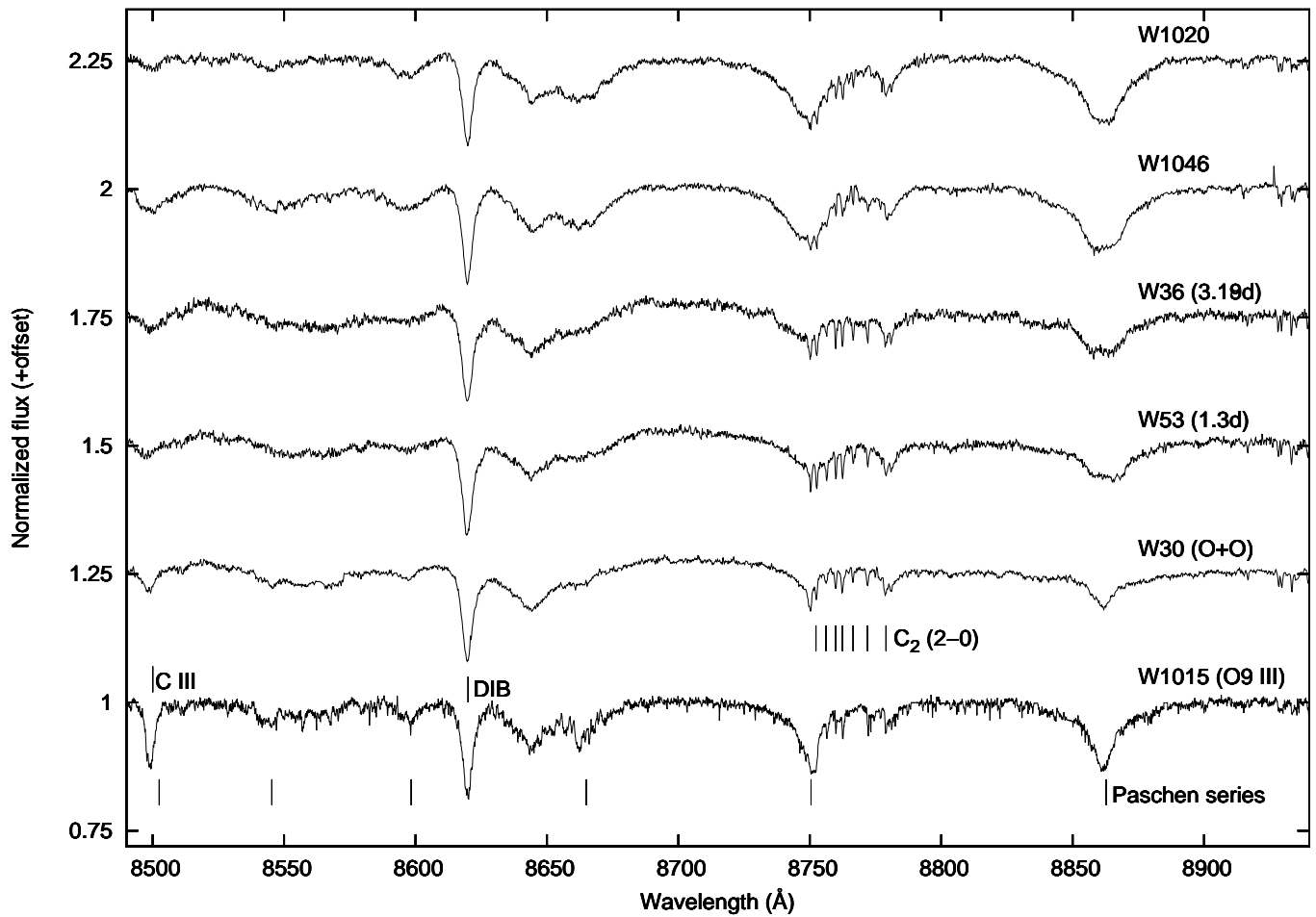}}
\caption{Comparison the $I-$band spectra of known and newly-identified
  candidate binaries with the O9~III star W1015. Wd1-30a
  is a strong X-ray source indicative of a colliding-wind binary,
  while Wd1-53 displays a 1.3-day periodic modulation in its
  light curve and double He\,{\sc i} lines, and Wd1-36 is a 3.18-day
  eclipsing binary.}
\label{fig:binaries}
\end{center}
\end{figure*}

While the spectral sequence shown in Figs. 2 and 3
provides a good general template for classification, many of the
faintest objects display a distinct I-band morphology, with weak, very
broad, and flat bottomed Pa-11 and Pa-12 lines, almost no trace of
Pa-13\dots15, and broad, weak C\,{\sc iii} 8500{\AA} (cf. Wd1-36; Fig. 2). The anomalous nature of
this population can be seen in Fig. 7, which plots
the width of the Pa-11 line against $I-$band magnitude for objects with
photometry from Clark et al. (\cite{clark05}). Cluster members show a linear
increase in the width of the Pa-11 line towards lower luminosity
classes, but many of the faintest objects display substantially
broader lines than this progression would suggest. In Fig. 8 we plot two representative examples  
of the broad lined population - W1020 and -1046 - against the O9~III classification
standard W1015, the binary systems Wd1-36 and -53 ($P\sim3.18$d and 
$\sim1.3$d respectively; Bonanos \cite{bonanos}) and the post-binary interaction  massive 
blue straggler Wd1-30a (O4-5Ia$^+$; Clark et al. \cite{clark19a}). It is clear from these data that the broad 
Paschen series lines of W1020 and -1046 are not a good match to the sharp bottomed photospheric 
profiles of W1015 and Wd1-30a (where the spectrum of the later is dominated by the early hypergiant component). 

Instead, given the strong morphological similarities that exist between W1020 and -1046
and the binaries Wd1-36 and -53, we suggest that many, if not \textit{all}, of the
observed broad-lined systems are SB2 binaries, with a composite spectrum
that includes contributions from a $\sim$O9~III primary and OB~III--V
secondary. Support for this conjecture is provided by:

\begin{itemize}
\item  The additional presence of comparable $I-$band
spectral morphologies in the photometrically identified 
binaries  Wd1-6a (Fig. 7) and W1021 ($P\sim2.20$d and $\sim4.43$d 
respectively; Bonanos  \cite{bonanos}); four such broad-lined systems 
are thus unambiguously  binary.

\item Despite our $I-$band
observations being poorly-suited to detecting radial velocity changes in
such broad-lined systems (Sect. 3.1), three stars - W1028, -1032, and -1061 - show radial
velocity shifts indicative of reflex binary motion ($\sigma_{\rm rv}\sim10-20$kms$^{-1}$; Clark et al.
in prep.). 

\item The appearance of anomalously broad and/or double troughed profiles in 
the photospheric H$\alpha$ and He\,{\sc i} 6678{\AA} and 7065{\AA} lines in our 
low-resolution FORS2/MXU spectroscopy (Fig. 6). Specifically W1046  clearly shows
a double-lined He\,{\sc i} 7065{\AA} profile, while W1014, W1028,
W1032 and W1044  all present H$\alpha$ and He\,{\sc i} lines
that are broad and shallow compared to both (apparently) single
stars (e.g. W1036; O9.5~Ib) and binaries in which the primary
is substantially more luminous than the secondary and hence dominates
the spectrum (e.g. W1065; B0~Ib).

\end{itemize}

As a consequence we provisionally apply generic classifications of 
O9-9.5III bin? and O+O? to  those stars itemised in footnotes 8 and 10,
respectively. The more luminous binaries Wd1-36 and -53a are likewise 
assigned an OB+OB classification while, as described previously, more 
precise classifications are possible for W1002, -1040, -1041 and -1055 (Table 2).

If this distinctive $I-$band morphology is indeed 
indicative of binarity then the implied binary fraction amongst 
the late-O population in Wd1 appears substantial. However, considerable 
caution must be exercised before accepting this assertion. An obvious 
 selection effect that must be allowed for is that binaries with an appreciable
contribution from a massive secondary will be more luminous than an
equivalent isolated star, and hence more likely to meet our
photometric criteria for selection. Moreover in a subset of  cases the broad-lined
morphology might reflect rapid rotation rather than binarity. As a 
consequence, confirming the binary nature of the majority of these broad-lined 
systems is clearly a priority. Unfortunately this is technically challenging 
since it will require high-resolution observations
targeting the He\,{\sc i} 6678{\AA} and 7065{\AA} photospheric lines,
as the reddening towards Wd1 precludes use of metal lines further bluewards. 
However the throughput with the appropriate VLT/FLAMES configuration yields
spectra of insufficient S/N to accomplish this goal in a reasonable timeframe (cf. Clark et al. in prep.).

\section{Discussion}\label{sec:discussion}

Our new observations yield a total of 69 new cluster members for Wd1, bring the total census to 
166 massive evolved stars. Understandably, given the fainter nature of the majority of targets (Fig. 1), we find that 
they are predominantly biased towards less evolved O stars of lower luminosity than the
cluster  OB supergiants previously classified (Table 1). We highlight that spectral classification for all objects has  been
 uniformly accomplished via a combination $I-$band and optical photometry, supplemented with $R-$band spectroscopy where 
available (Sect. 3 and Negueruela et al. \cite{neg10}); as a consequence we consider this process to be internally robust.

 A particular problem with  the quantitative interpretation of these data 
is that Wd1 suffers from  heavy differential reddening and the   form of 
the extinction law is 
ill-constrained, which in turn leads to a large range of possible 
bolometric luminosities for individual cluster  members (Clark et al. 
\cite{clark19a}).  As a result we refrain from  updating the 
semi-empirical cluster  HR diagram introduced by Negueruela et al. 
(\cite{neg10})  at this time; instead summarising the data via a 
colour/magnitude plot (Fig. 9 and Sect. 4.1).

Nevertheless a  number of features are obvious from the current analysis:
\begin{itemize}
\item The presence of a smooth progression in spectral morphology/stellar classification of evolved stars from late-O giants through to  early-mid B supergiants and ultimately late B hypergiants, with the supergiant population dominated by spectral types
O9-B1 (60 examples). 
\item An unprecedentedly rich population of both hot and cool hypergiants with, uniquely, spectral types ranging from O4-5 Ia $^+$ (Wd1-30a) 
to F8 Ia$^+$ (Wd1-8a) and, arguably,  M5 Ia (Wd1-20). 
\item A population of WRs exhibiting a diverse array of spectral sub-types.
\item A large number of spectroscopic binary candidates, including a number of interacting/post-interaction systems. 
\end{itemize}

We discuss the implications of these findings in more detail below, in the context of both stellar evolution and the bulk properties of Wd1.

\subsection{Stellar content}

The extreme mass inferred for Wd1 (Clark et al. \cite{clark05}, Lim et al. \cite{lim}, Andersen et al. \cite{andersen}) implies that even very rapid/rare phases of massive stellar evolution may be represented. Conversely at an apparent age of $\sim$5Myr, we would expect stars of initial mass $M_{\rm init}\gtrsim60M_{\odot}$ to have been lost to core-collapse\footnote{Groh et al. (\cite{groh13}) predict that for an age of 4Myr non-rotating (rotating) stars of $\sim60M_{\odot}$ 
($M_{\rm init}\sim85M_{\odot}$) will currently be underging core-collapse and at 5Myr $M_{\rm init}\sim40M_{\odot}$ 
($\sim60M_{\odot}$).}; as a consequence it is also of interest to review the stellar subtypes that are not present. Considering first the least evolved members and, as might be expected, very early O2-4 I-III stars  - found within the youngest ($\lesssim2$Myr) clusters such as R136 and NGC3603; Crowther et al. \cite{crowther16}, Melena et al. \cite{melena}, Roman-Lopes et al. \cite{roman}) - are absent from Wd1. Instead, the earliest stars present are the O4-5 Ia$^+$ and O7-8 Ia$^+$ hypergiants Wd1-30a and Wd1-27, which appear to be the products of binary evolution  (Sect. 4.4; Clark et al. \cite{clark19a}). Discounting these stars on this basis, the earliest `canonical' supergiants are of spectral type $\sim$O9Ib...Iab (e.g. Wd1-15 and -17). 

The presence of a rich population of late-O to mid-B supergiants has long been recognised 
(Clark et al. \cite{clark05}, Negueruela  et al. \cite{neg10}). 
In contrast to such earlier spectroscopic studies, our observations include
large numbers of lower luminosity objects in Wd1, with 
target selection based on photometry consistent with heavily-reddened
OB stars. These  criteria would not distinguish between
O9~III stars evolving to become such luminous OB supergiants - which are seen in large numbers - 
and early B0.5--2 Ib...II stars 
from an older population of stars with lower initial masses currently 
evolving to the RSG phase. Such stars would be readily distinguished from the O9~III  population by the presence
of strong He\,{\sc i} (and possibly N\,{\sc i}) lines and the absence of C\,{\sc iii}.
Critically they are entirely absent from  our survey;  the only B-type
objects are super-/hypergiants of sufficient luminosity that they are consistent with the wider evolved stellar  population of
Wd1 (Sect. 4.2).

 Examples of every transitional evolutionary phase between H-rich OB supergiants and H-depleted  WRs are found within Wd1 (e.g. BHG, 
 LBV, sgB[e], YHG, and  RSG). Surprisingly, only one LBV  is identified (Wd1-243; Clark \& Negueruela \cite{clark04}),
 compared to large numbers of  confirmed and candidate systems within, for example, the Quintuplet (Clark et al. \cite{clark18b})\footnote{We 
note that Gvaramadze (\cite{gv}) suggest that the LBV MN44 is a runaway from Wd1.}. 
Intriguingly, Wd1-243 also has a rather moderate mass-loss rate (Ritchie et al. \cite{ritchie09b}, Fenech et al. \cite{fenech18})
 in comparison to other (candidate) LBVs such as P Cygni (Najarro et al. \cite{najarro}), AG Car (Groh et al. \cite{groh09}) and HDE~316285 (Hillier et al. \cite{hillier98a}) where, unlike Wd1-243, the wind densities are sufficient to drive the Paschen series into emission. One might appeal to the comparatively low temerature of Wd1-243 to explain this discrepancy; however the Paschen series is seen in emission in the YHG IRAS 18457-0604, which appears to be of comparable spectral type to Wd1-243 (Clark et al. \cite{clark14a}). Expanding on this point, and it is noteworthy that the population of YHGs within Wd1 also lack the pronounced emission line spectra of other examples, such as IRAS 18457-0604 and IRC +10 420 (Clark et al. \cite{clark14a}). Apparently comparatively low mass loss rates are a feature of the the current population of LBVs and YHGs within Wd1, although the presence of extended mm-/radio nebulae associated with a number of these stars is indicative of extensive mass loss in the recent past (Dougherty et al. \cite{dougherty}, Andrews et al. \cite{andrews}, Fenech et al. \cite{fenech18}).

Finally we turn to the WR content of Wd1. A wide range of WN sub-types are present, although  the earliest (WN2-4) examples are not represented (Table 4). No WN5-9ha stars are observed; consistent with the expectation that they descend from exceptionally massive stars (e.g. Lohr et al. \cite{lohr18a}, Schnurr et al. \cite{schnurr}, Bonanos et al. \cite{20a}).
Three late WN9-11h stars are identified (Wd1-5, -13 and -44); however all are potentially the products of binary interaction (Ritchie et al.  \cite{ritchie10}, Clark et al. \cite{clark14b}, in prep.). In contrast to the WN stars, the WC stars are uniformly of late (WC8-9) spectral subtype; neither transitional WN/WC nor WCE stars are identified; we return to this in Sect. 4.3.
Lastly, Groh et al. (\cite{groh13}) predict that non-rotating (rotating) stars of $M_{\rm init}\gtrsim 50M_{\odot}$ ($\gtrsim 32M_{\odot}$) present as WRs of WO subtype immediately  prior to core-collapse. Once again no examples  are found within Wd1; however with predicted optical magnitudes 
$\gtrsim1.5$mag fainter than those of O9-9.5 III stars, this is perhaps unsurprising (Groh et al. \cite{groh13}, Martins \& Plez 
\cite{mp06}). 

\begin{figure*}
\begin{center}
\resizebox{\hsize}{!}{\includegraphics{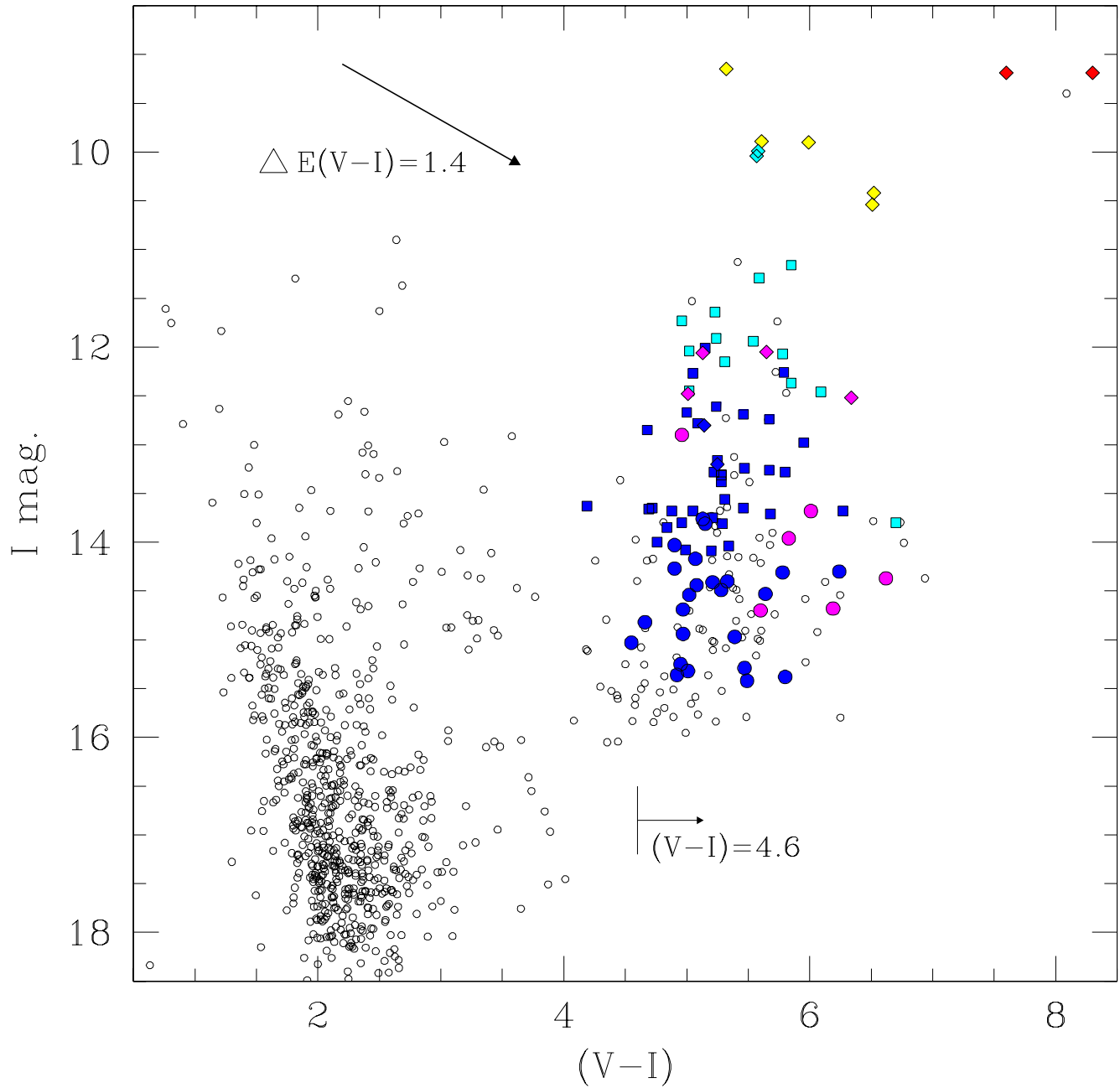}}
\caption{Colour magnitude plot of the 5'$\times$5' field-of-view centred on Wd1 utilising the photometry of Clark et al. 
(\cite{clark05}). Cluster members with both spectral types and luminosity classes are denoted as follows: RSGs and YHGs - red and yellow diamonds; B5-9 Ia$^+$ - cyan diamonds; B1-4 Ia,Iab,Ib - cyan squares; O9-B0.5 Ia,Iab,Ib - blue squares; 
O9-B0.5 II,III - blue circles; early BHG/WNVLh - purple diamonds; WN5-8 \& WC8-9 - purple circles. The vector shows the effect of differential reddening across the cluster (cf. Negueruela et al. \cite{neg10}) assuming an `off-the-peg' Cardelli et al. 
(\cite{card}) prescription. A colour cut for cluster membership of 
$(V-I)\sim4.6$ as suggested by the OB (super)giants (Sect. 4.1) is also 
indicated.}
\label{fig:standards}
\end{center}
\end{figure*}

  We may summarise these results graphically via the updated 
 colour/magnitude diagram given in Fig. 9, constructed with the dataset 
 presented in Clark et al. (\cite{clark05}); inevitably  this excludes the 
 62 stars with spectral classifications but lacking suitable photometry (Table 2). 
 Nevertheless, remaining cluster members are clearly delineated from field 
 stars by their excessive reddening, with the detection threshold set by 
 the $V-$band magnitude. Stars assigned  both a spectral types and 
 luminosity class are further indicated; unfortunately  a subset of 
 binaries  cannot be designated on this basis (Wd1-9, -36, -53a and those 
 in footnote 10).  
 The progression in both magnitude and colour with spectral classification 
 is broadly as expected with, for example, the cool super/hypergiants being 
 systematically redder and brighter than the OB (super-)giants. The 
 breadth of the region occupied
 by cluster members in the diagram is indicative of significant 
 differential reddening, with Negueruela et al (\cite{neg10}) 
 finding ${\Delta}E(V-I)\sim5.0-6.4$mag and  a mean of $\sim5.5\pm0.3$mag 
 from cluster blue supergiants.  Utilising the colours of OB stars 
presented by Wegner (\cite{wegner}) this suggests a colour cut for cluster 
members of $(V-I)\sim4.6$; as shown in Fig. 9, with the exception of a 
single outlier this indeed serves as a suitable discriminant. 
 Uncertainties in spectral classification and the effect of binarity 
 likewise contribute to the overlapping  regions occupied by, for example,  
 O9-B0.5 stars of luminosity classes  I and II...III (cf. Sect. 3.4.1 and 
 3.5). Finally the diagram reveals  the residual incompleteness of our 
 current spectroscopic survey. This is particularly prevalent amongst the 
 fainter cohort of members, although this shortcoming also extends to 
 stars within the cluster core with luminosities suggestive of a 
 supergiant nature, but which still await spectral classification (e.g. 
Wd1-12b, -39a and 40a) and, in some cases, photometric observations.

\subsection{Bulk properties of Westerlund~1}

\subsubsection{Cluster age}

 Previous studies attempting to determine the age of Wd1 have focused on both high- and low-mass 
stellar components. Placement of the OB super-/hypergiant population on a semi-empirical HR-diagram 
suggests a high degree of coevality for this  cohort, with a comparison to evolutionary tracks 
implying an age of $\sim5$Myr (Negueruela et al.  \cite{neg10}, Meynet \& Maeder \cite{meynet03}, 
Ekstr\"{o}m et  al. \cite{ekstrom}). 
Similar conclusions were  drawn from a
 proper-motion selected photometric study by Kudryavtseva et al.  
(\cite{kud}), who were able to  
demonstrate that the form of the pre-MS transition in Wd1 is consistent with a co-eval population at 
an age of 5Myr with a comparatively small  uncertainty ($<0.4$Myr).
However age determinations via both techniques rely on the applicability of isochrones derived  from 
evolutionary codes for single stars and are further   hampered by uncertainties in distance and 
extinction measurements (cf. Clark et al. \cite{clark19a}), leading to e.g. a  degeneracy in the 
placement of 3-5Myr pre-MS isochrones 
(cf. Andersen et al. \cite{andersen}, Hosek et al. \cite{hosek}).

 A unique feature of Wd1 is the simultaneous presence of both WRs and 
 RSGs; under the  assumption of single star evolution for 
both cohorts this has led previous 
 studies to infer a relatively narrow age range of $\sim4.5-5$Myr (Clark 
 et al. \cite{clark05}, Crowther et al. \cite{crowther06}). Recent 
 research has again emphasised the role that the presence and nature  
 of cool supergiants may play in constraining the ages of young massive 
 clusters, especially if  the placement of the main-sequence  is uncertain 
 due to the effects of binary interaction (e.g. Britavskiy et al. 
 \cite{brit}, Beasor et al. \cite{beasor}). However current evolutionary 
 predictions by different groups fail to converge on a consistent set of 
physical properties for this phase (cf. comparison in Britavskiy et al. 
 \cite{brit}; see also Davies et al. \cite{davies18} for a comparison to 
 observations); potentially complicating the employment of such a diagnostic.

 Unfortunately a combination of saturation and blending of the Wd1 RSGs at 
 IR wavelengths precludes accurate determinations of their luminosities 
 for comparison to such predictions. Clark et al. (\cite{clark05}) suggest 
 the cluster YHGs  are particularly luminous ($L_{\rm 
 bol}/L_{\odot}\gtrsim5.7$); if such stars evolve at essentially constant 
 bolometric luminosity they will yield RSGs that are more luminous than 
 those inferred to exist  within the Magellanic Clouds (Davies et al. 
 \cite{davies18}), although nominally  consistent with the predictions of 
Ekstr\"{o}m et al. (\cite{ekstrom}) for non-rotating stars. As a consequence, 
quantitative determination of physical 
 properties of the RSG cohort of Wd1 may prove to be particularly valuable 
 in both constraining the cluster  age and, more widely, in validating 
evolutionary predictions for this phase of stellar evolution.

 Nevertheless, given that the uncertain extinction law towards Wd1 greatly 
 complicates tailored quantitative analyses for individual hot and cool 
 stars (cf. Clark et al. \cite{clark14b}, \cite{clark19a})  we defer the  
 assembly  of an empirical HR diagram - and consequent comparison of the 
 cluster population to theoretical isochrones - until this issue is 
 resolved. As such we are currently unable to improve on the quantitative 
 age estimate of Negueruela et al.  (\cite{neg10}). Notwithstanding this, 
 the newly classified cohort of early-type stars appears qualitatively 
 consistent with  the assertion that the (massive) stellar population of 
 Wd1 is essentially co-eval. Specifically, the super-/hypergiant evolutionary sequence 
described by Negueruela et al. (\cite{neg10}) can now be extended to lower-luminosity 
giants  of systematically earlier (O9-9.5) spectral type,  
yielding a remarkably homogeneous population of OB stars within Wd1. 
In the (current) absence of an identifiable population of core 
H-burning stars of  luminosity class V, the sole examples of O stars of even 
earlier ($<$O9) spectral type appear to be binary products, while no B supergiants 
descending from stars of lower initial masses have been distinguished. 
We therefore conclude that, subject to completeness (Sect. 4.1), our
 current stellar census provides no qualitative evidence for younger or older stellar populations within Wd1.

\subsubsection{Stellar and cluster masses}

 Utilising a combination of stellar evolution and non-LTE model-atmosphere codes, Groh et al. 
(\cite{groh14}) and Martins \& Palacios (\cite{mp}) calculated synthetic spectra for 
single non-rotating  massive stars, from which empirical spectral classifications can be derived for comparison to real data. Both 
sets of calculations  imply that stars with  $M_{\rm init}\sim60M_{\odot}$ fail to yield the early-mid B supergiants  present within 
Wd1; indeed at an age  of $>4$Myr such stars should already have been lost to SNe (footnote 11). Conversely, the late-O giants 
identified in this work  do not form under $M_{\rm init}\leq20M_{\odot}$ pathways. This leaves a  $20M_{\odot} < M_{\rm 
init} <50M_{\odot}$ window which appears to yield OB stars of the spectral type and luminosity class found within Wd1, with  the WR cohort deriving from the upper reaches of this range. 
The eclipsing binary Wd1-13 provides direct support for this supposition, with Ritchie et al. 
(\cite{ritchie10}) deriving current dynamical masses of $\sim23.2^{+3.3}_{-3.0}M_{\odot}$ 
and $\sim35.4^{+5.0}_{-4.6}M_{\odot}$ for the B0.5 Ia$^+$/WNVLh and OB supergiant components, respectively. Under 
highly non-conservative late-Case A/Case B mass transfer, a pre-interaction $M_{\rm init}\sim40M_{\odot}$ is inferred for the primary. Similarly high progenitor masses are suggested by quantitative analyses of Wd1-5, -27 and -30a (Clark et al. \cite{clark14b}, \cite{clark19a}). Although subject to systematic uncertainties in both cluster distance and and extinction law the semi-empirical HR diagram of Negueruela et al. (\cite{neg10}) is also fully consistent with this hypothesis.

Outside of Wd1, few dynamical mass-estimates for late-O/early-B stars of luminosity class I-III are available; being limited to
CC Cas (O8.5 III, $35.4\pm5M_{\odot}$; Gies \cite{gies}), the secondary in the detatched binary HD 166734 (O9 
I(f), $\sim33.5_{-3.7}^{+4.6}M_{\odot}$; Mahy et al. \cite{mahy}) and the evolved, post-interaction primary in Cyg OB2-B17
(O9 Iaf, $45\pm4M_{\odot}$; Stroud et al. \cite{stroud}). Nevertheless these are consonant with both values derived from Wd1 and 
also the  the mass range inferred for field B0-3 Ia stars ($\sim25-40M_{\odot}$) by  Crowther et al. (\cite{crowther06b}). 

Following from the above we may attempt to infer an integrated mass for Wd1 from the massive star census, 
assuming a Maschberger (\cite{masch}) IMF with standard parameters,
including an upper-mass slope of 2.3 (ie. Salpeter).  If we take a highly  conservative limit that the 166 spectrally classified 
cluster members span a  range of 20--50 M$_\odot$
then the {\em current} mass of Wd 1 (ie. the mass of stars between 0.1 and 50
M$_\odot$) is $\sim 6 \times 10^4$ M$_\odot$ (with a $\sim5$ per cent error due to
the relatively low-number of massive stars). 

However, with a rather more realistic mass  range of 25--50 M$_\odot$,  the
{\em current} mass of Wd 1 is $\sim 9 \times 10^4$ M$_\odot$ (again with a $\sim5$
per cent error).  This significant increase is due to the high-power of the upper mass
slope meaning that shifting the lower mass limit from 20 to 25 M$_\odot$ increases the
total mass by 50 per cent,  showing the crucial importance of accurately
determining the lower mass limit. Conversely, the upper mass limit is much less important; for example if 
we were to specify a range of masses of 20--45 M$_\odot$, the estimated integrated mass would increase by less than 10 per 
cent.

We refrain from specifying a contribution from higher
mass stars since it would require an extrapolation of the mass function to stars already lost to core-collapse
and likely comprises a comparatively minor contribution to the integrated cluster mass. Moreover, we caution that if 
the complete initial mass function departs from a Salpeter form - for example if it is top-heavy -  the above 
masses would be over-estimates.

\subsection{Comparison to other young massive clusters}

At this juncture it is instructive to compare the stellar population of Wd1 to other young massive clusters. Both 
NGC3603 and R136 are sufficiently young ($\sim1-2$Myr; Melena et al. \cite{melena}, Crowther et al. \cite{crowther16})  
that their demographics  are very different from Wd1 and so we do not consider them further here. Of more relevance are the three
clusters in the central molecular zone of the Galaxy. Both the Arches and Quintuplet are expected to be younger than Wd1 ($\sim2-3.3$Myr and $\sim3-3.6$Myr respectively; Clark et al. \cite{clark18a}, \cite{clark18b}, \cite{clark19b}); however, the situation is less clear cut for the Galactic Centre cluster, with an age of  $6\pm2$Myr typically quoted (Paumard et al. \cite{paumard})\footnote{Such a large age spread results from uncertainties in both population synthesis and the quantitative modeling of cluster members (Paumard et al. \cite{paumard}, Martins et al. \cite{martins07}). However at most three RSGs may be physically associated with the Galactic Centre cluster   (Blum et al. \cite{blum}) implying a much more extreme WR/(YHG+RSG) ratio than found for Wd1 (33:3 versus 24:10); potentially indicative of a younger age (Davies et al. \cite{davies09}). Such  an assertion is further supported by the rather high dynamical masses inferred for the components of the binary GCIRS 16SW ($\sim2\times50M_{\odot}$; Martins et al.\cite{martins06}).}.

Once differing ages are taken into account, similarities are apparent  between the evolutionary  sequences of H-rich cluster members of the Arches,  Quintuplet and Wd1 (with the Galactic Centre population insufficiently sampled for comparison). Specifically, stars in the three 
clusters demonstrate a smooth progression in spectral morphologies extending from OB supergiants  (O4-6 Ia, O7-B0 Ia and O9-B4 Ia respectively) through to hypergiants (O4-8 Ia$^+$, B0-3 Ia$^+$ and B5-9 Ia$^+$). While absent from the Arches due to its youth,  more advanced evolutionary phases, such as LBVs and the related WN9-11h stars, are found within the Quintuplet, with  the YHGs and RSGs playing an analogous  role in Wd1. Moreover, both Wd1 and the Quintuplet  contain a cohort of stars with anomalously early spectral types\footnote{qF256 (WN8-9ha), -274 (WN8-9ha) and -406 (O7-8 Ia$^+$) and  LHO01 (O7-8 Ia$^+$/WNLha) and -54 (O7-8 Ia$^+$) within the Quintuplet and Wd1-5 (WN10-11h/B0.5 Ia$^+$), -13 (B0.5 Ia$^+$), -27 (O7-8 Ia$^+$), -30a (O4-5 Ia$^+$), -44 (WN9h),  and W1049 (B1-2Ia$^+$).}, of which a number would fit seemlessly into the Arches cluster; potentially indicative of binary interaction and rejuventation (Sect. 4.4 and Clark et al.  \cite{clark18a}, \cite{clark18b}).

However the WR cohorts of the four clusters look very different from one another (Table 4). The Arches appears too young for any 
H-free examples to be present, with a population solely confined to WN7-9ha stars (Clark et al. \cite{clark19b}). Conversely, WN sub-types appear surprisingly under-represented  within the Quintuplet, which contains just three WN9-11h, two WNLha (footnote 14) and a single WN6 star -  despite Groh et al. (\cite{groh14}) predicting  a WNE phase lasting  $\sim3\times$ longer than the WNL phase. Instead the cluster is dominated by WC8-9 stars (Table 4), many of which appear to be binaries identifiable by excess IR emission from hot dust (Clark et al. \cite{clark18b} and refs. therein); again none of the WCE stars anticipated by Groh et al. (\cite{groh14}) are present. In stark constrast, the  WN:WC  star ratio is reversed for Wd1, with all spectral 
sub-types from WN5 to WN11h represented, albeit with an apparent (weak) bias towards earlier (WN5-7) stars. 

The WN:WC ratio for the Galactic Centre cluster is more evenly balanced although, unlike Wd1, the WN population appears biased towards later spectral sub-types (Table 4). As with the Quintuplet and Wd1, the WC population is dominated by WC8-9 sub-types, with the exception of a single WC5/6 star. Based on spectral morphologies, Martins et al. (\cite{martins07}) suggest an evolutionary sequence of:\newline

WN9h$\rightarrow$WN8$\rightarrow$WN/WC$\rightarrow$WC8-9\newline

for stars within this  cluster. No such simple scheme can be constructed for Wd1 due to an absence of transitional WN/WC stars 
and the range of WN spectral sub-types observed. Similarly the almost complete lack of H-free WN stars within the Quintuplet complicates interpretation of evolution from the  BHG/LBV/WN9-11h to WC phase at very high initial masses ($\gtrsim60M_{\odot}$;
Clark et al. \cite{clark18b});   we return to these observational findings in Sect. 4.4.

 We may also extend this comparison beyond core-collapse. One of the defining features of Wd1 is the presence of the  magnetar CXO 
 J164710.2-455216 (Muno et al. \cite{muno06a}). Intriguingly, a magnetar has also been associated with the Galactic Centre cluster
 (SGR J1745-29; Kennea et al. \cite{kennea}, Mori et al. \cite{mori}), while PSR J1746-2850 - either a high $B-$field  pulsar or 
 transient magnetar - is located in close proximity  to the Quintuplet ($\sim5pc$; Deneva et al. \cite{deneva}, Dexter et al. 
 \cite{dexter}) although, given the population of  isolated massive stars in the central molecular zone, an unambiguous  physical 
 association has yet to be proven. Finally, in terms of stellar content - and hence age - the host cluster of SGR1806-20 appears to be 
 directly comparable to the Quintuplet (Figer et al. \cite{figer05}, Bibby et al. \cite{bibby}). These observations strongly suggest 
that  clusters in the $\sim3-6$Myr age window appear efficient sites for their 
 formation, despite a naive expectation that under a single star evolutionary channel such massive progenitors should instead form 
 black holes. In turn this implies that either the  physical process by which neutron stars form from 
massive 
progenitors favours the production of magnetars (cf. Clark et al. \cite{clark14a}) or that (counter-intuitively) a large 
fraction of neutron stars are born as 
magnetars; a conclusion recently reached  by Beniamini et al. (\cite{ben}). 

\subsection{Binarity and evolutionary channels}

Two key findings of this work are the unprecedented range of hypergiants within Wd1 - not reproducible for a co-eval stellar      
 population under single star evolution (e.g. Ekstr\"{o}m et al. \cite{ekstrom}, Brott et al. \cite{brott}) - and the apparently rich 
 binary population amongst the earliest O stars present. Specifically, of the 27 O9-9.5 III stars within Wd1 (Table 2), one is a 
 confirmed photometric binary (W1021) and   13 have anomalously broad Paschen lines apparently indicative of binarity (of which three 
 are RV variable; Sect. 3.4.3 \& 3.5 and  footnote 8). Moreover, a further 10 objects have Paschen lines sufficiently broad to 
preclude precise classification but otherwise appear observationally  consistent with an O+O designation (footnote 10). This appears indicative of a remarkably high binary fraction, although we caution that with both components of comparable luminosity - as shown by their composite spectra - such systems will be systematically brighter than their single siblings, introducing a 
potentially significant bias into their identification. Irrespective of this,  we also identify similar candidate spectroscopic binaries amongst brighter, more evolved stars  (footnote 9), while we are able to confirm that a number of photometric binaries from Bonanos 
(\cite{bonanos}) are indeed luminous OB cluster members (e.g. Wd-36, -53, and  W1048).

These results are consonant with previous observations which imply a large number of massive binaries within Wd1. Examples  include (i) a binary  fraction of $\gtrsim70$\% inferred for WR stars via IR and X-ray observations (Crowther et al. \cite{crowther06}, Clark et al. \cite{clark08}, Clark et al. \cite{clark19b}) and (ii) the embodiment of very short-lived phases of binary interaction such as  Wd1-9, which is thought to be undergoing rapid case-A mass transfer (Clark et al. \cite{clark13}). 

 Given the wealth of observational diagnostics, we defer a quantitative analysis of the full binary population of Wd1 to a future paper, where we present a synthesis of multiple datasets, including results from our full RV spectroscopic survey  (Clark et al. in prep.).  Nevertheless, just considering the  data presented  here provides compelling observational evidence for the role that binary-driven evolution plays within Wd1. 

Following Negueruela et al. (\cite{neg10}) we may presume  a single star evolutionary channel broadly progressing: \newline

late-O giants $\rightarrow$ early-B supergiants $\rightarrow$ mid to late-B hypergiants $\rightarrow$  YHG/RSGs\newline

 before looping back bluewards to H-depleted WRs, potentially via an LBV phase (cf. Wd1-243). However neither the mid-O nor the early-B hypergiants within Wd1 are  consistent with such a scenario and hence  we must suppose two futher, binary-modulated channels to yield these stars (cf. Clark et al. \cite{clark19a}).

 The first is uncontroversial and presumably occurs as the primary evolves towards the supergiant phase and fills its Roche lobe, with  the resultant  binary-driven mass stripping yielding  undermassive but overluminous and chemically peculiar WNLh/BHG stars such as Wd1-5, -13, -44 and, presumably, W1049 (Petrovic et al. \cite{petrovic}, Ritchie et al. \cite{ritchie10}, Clark et al. \cite{clark14b}). Secondly one might suppose that such  binary interaction may also lead to  the secondary accreting significant quantities of mass or, in extreme cases, merger. In this case one would expect a very luminous and massive blue straggler to form; examples being 
Wd1-27 and -30a (de Mink et al. \cite{demink14}, Schneider et al. \cite{schneider14}, Clark et al. \cite{clark19a}).
Such a channel is more controversial since it requires the secondary to be able to accrete large quantities of mass without spinning it  up to critical rotation via the transfer of angular momentum (which would quickly halt accretion; Petrovic et al. \cite{petrovic}, 
de Mink et al. \cite{demink14}); physical mechanisms that might facilitate this  include angular momentum loss via an accretion disc or tidal interaction between both components.

Unfortunately, the relative prevalence or weighting  of both channels is currently unclear for a number of reasons. Evolutionary codes including all relevant physical processes have yet to be constructed, with the efficiency of the accretion of mass and angular momentum essentially treated as a free parameter at this point. This uncertainty also means that subsequent, post-interaction evolution is opaque; given the wide range of WN sub-types present within Wd1 it would appear highly  likely that some result from a binary evolutionary channel (cf. Clark et al. \cite{clark19c}), but we lack theoretical predictions to validate this conclusion (cf. Groh et al. \cite{groh14}). 
Observationally, one might expect that in many  realisations of mass-stripping the secondary dominates the emergent, post-interaction spectrum, preventing  identification and characterisation  of the primary (cf. G\"{o}tberg et al. \cite{strip}). 

Focusing specifially on Wd1 and since the secondary in Wd1-13 superficially resembles other cluster supergiants (Ritchie et al. \cite{ritchie10}), one might suppose that a number of apparently single OB stars within Wd1 are also in a post-interaction phase.
However extensive and systematic quantitative model-atmosphere analysis will be required to identify such binary products via their rapid rotation and/or anomalous chemical 
abundances. Indeed we caution that while the mass ratio has reversed in Wd1-13, we cannot empirically constrain the quantity of mass the secondary has accreted at this time, and hence whether the system  evolved via (quasi-)conservative mass-transfer.

As a consequence we are simply left with the conclusion that both  channels must operate in parallel in Wd1, with Wd1-27 and -30a indicating that mass-transfer may be very efficient (Clark et al. \cite{clark19a}). Conversely, the extreme mass loss rate exhibited  by the sgB[e] star and interacting binary Wd1-9 reveals that  in certain instances much of the mass stripped from the primary  is lost from the system rather than accreted by the secondary (Clark et al. \cite{clark13}, Fenech et al. \cite{fenech17}, \cite{fenech18}).

\section{Concluding remarks and future prospects}

In this work we present classifications of  a further 69 members of Wd1, producing a current  census 
of 166 massive, evolved stars. As expected, given the photometric 
 selection criteria adopted, the majority of these are  late-O stars of 
 luminosity type I-III, which smoothly extend the  morphological sequence 
 of hot and cool super-/hypergiants  identified by Negueruela et al. 
 (\cite{neg10}; Sect. 4.4) to higher temperature and less evolved objects.  While a 
 handfull of B-type supergiants and two new hypergiants have also been 
 identified, no lower luminosity B stars - which would be indicative of an 
 older stellar population - have been discovered. Likewise we find no 
 evidence for a younger, more massive population.  We conclude that on 
 this basis our current stellar census is consistent with the hypothesis 
 that Wd1 is co-eval.

Unfortunately, given the limitations of these data and uncertainties in 
both the distance and extinction towards Wd1 it is premature to construct 
an HR diagram for the cluster and so we may not improve on previous    
 quantitative age estimates  via comparison to theoretical isochrones, or 
provide initial masses for individual stars. However comparison to the the 
stellar population of the $\sim3-3.6$Myr old  Quintuplet (Clark et al. 
\cite{clark18b}) reveals systemtic differences,  most noticably in the 
spectral type distributions of OB super-/hypergiants,  that indicate that 
Wd1 is older. Indeed the updated census of massive stars presented here is 
 consistent  with previous age estimates of $\sim5$Myr derived via a 
number of different methodologies (Sect. 4.2.1).

Moreover,  qualitative comparison of the observations to the 
single star synthetic spectra generated by Martins \& Palacios (\cite{mp}) 
 suggests that, collectively, the cluster members sampled  evolved from  
 $20M_{\odot} < M_{\rm init} <50M_{\odot}$ progenitors (Sect. 4.2), with 
 more massive stars having already 
been lost to SNe  (Groh et al. \cite{groh13}).  This would in turn imply 
that we may expect  to see an evolutionary turn-off within  Wd1  around 
O7-8 V, corresponding   to stellar masses of $\sim25M_{\odot}$ (Martins \& 
Palacios \cite{mp}, Clark et al. \cite{clark19a}). 

Based on spectral morphology  and classification we may  construct an 
apparent evolutionary sequence  for H-rich single stars within Wd1 up to 
the hypergiant phase that is  analogous to that inferred for the 
Quintuplet (Sect. 4.3 \& 4.4).  However, this  progression cannot be 
extended to the H-free WR cohort  for either cluster, with that of Wd1 
found to be unexpectedy diverse (Table 4).  Moreover a population of massive blue 
stragglers, with properties inconsistent  with these evolutionary 
pathways, may also be identified within both  aggregates (footnote 14). 
The simplest explanation for these phenomena is  the onset of binary 
interaction as stars evolve beyond luminosity class V.

Consonant  with both this hypothesis and previous observational studies, our data suggests a large binary fraction for Wd1 for stars at every evolutionary stage. These comprise both pre-interaction (W1021), interacting (Wd1-9) and post-interaction systems (Wd1-5, -13, -27, -30a and WR A and B). The properties of the latter cohort are indicative of two distinct but related evolutionary channels producing low-mass stripped primaries and  {\em potentially} rejuvenated, mass-gaining secondaries -  the extreme mass-loss associated with the sgB[e] star Wd1-9 indicating that in some instances mass lost from the primary is ejected from the system rather than accreted. Unfortunately our current observations only identify the most extreme examples of massive blue stragglers resulting from the latter pathway. Determining  the relative prevalence of both evolutionary channels will require quantitative analysis of the full cohort of OB
super-/hypergiants in order to identify the subset which are potential binary products via their  rapid rotation, anomalous surface abundances and/or mass luminosity ratios.

 Allowing for both incompleteness and the presence of massive binary 
companions, a conservative estimate 
suggests that Wd1 currently contains $>166$ stars with $M_{init} \gtrsim20M_{\odot}$.
Assuming a standard Maschberger IMF this would imply a 
rather extreme integrated cluster
 mass of $\sim 9\times10^4M_{\odot}$ (Sect. 4.2).

As a consequence of  their large integrated masses, the rich stellar populations of Wd1 and the  Arches and Quintuplet 
 clusters provide a unique opportunity to study the  post-main sequence evolution 
 of stars with $M_{\rm init} \sim20M_{\odot}$ to 
 $\gtrsim120M_{\odot}$ in the $\sim2-5$Myr window  (Clark et al. 
\cite{clark18b}, \cite{clark19b}).  Inclusion of clusters 
such as NGC3603 or Westerlund~2  would extend this 
investigation to relatively unevolved very massive  stars ($\sim1$Myr and $\gtrsim80M_{\odot}$; Melena et al. \cite{melena}, Zeidler 
et al. \cite{zeidler}, Bonanos et al. \cite{20a},  Schnurr et al. 
\cite{schnurr}). Conversely, older aggregates such as  Berkeley 51 \& 55  and 
the RSG dominated clusters at the base of the  Scutum-Crux  arm 
($\sim10-50$Myr; Lohr et al. \cite{lohr18b}, Davies et al. \cite{davies}, 
Clark et al. \cite{clark09})  permit the lifecycle of stars of 
initial masses down to $\sim8M_{\odot}$ - delineating the onset of  
core-collapse - to be explored. Of particular interest is the prospect 
of characterising the properties of  stellar subtypes  which have hitherto been 
poorly studied due to the brevity  of the evolutionary phase; a 
possibility exmplified by the uniquely  rich population of  hot and cool 
hypergiants within Wd1.

One may also extend such efforts to core-collapse and beyond. Groh et al. (\cite{groh13}) predict that non-rotating (rotating) stars of $M_{\rm init}\sim32M_{\odot} (28M_{\odot})$ will undergo core-collapse at an age of 5.81Myr (7.92)Myr; thus depending on the admixture one would expect all 166 stars within this census to be lost within $\lesssim3$Myr, implying a time-averaged rate of core-collapse of one every $\lesssim$18,000yr over this period (comparable to the rate inferred by Muno et al. \cite{muno06b}). As a consequence one might anticipate that  the immediate progenitors of such a process will be present at this time and, if  identifiable, ameanable to quantitative analysis. 

This has clear implications for the nature of the resultant relativistic remnant; particularly interesting since clusters such as Wd1 appear to function as highly efficient factories for the production of magnetars (Sect. 4.4). Intriguingly the anticipated magnetic decay timescale for magnetars ($\sim10^4$yr; Beniamini et al. \cite{ben} and refs. therein) is directly comparable to the SN rate within Wd1, suggesting that many core-collapse events must produce such objects. 
This raises the prospect of delineating the formation channel for magnetars (Clark et al. \cite{clark14b}) - of particular interest given that they  have been hypothesised to power superluminous SNe (Thompson et al. \cite{thompson}, Woosley \cite{woosley}, Kasen \& Bildsten \cite{kasen}). Moreover it immediately raises the question of how black holes 
of masses such as those found in high-mass X-ray binaries (e.g. Cyg X-1 with $M_{\rm BH}\sim14.8\pm1.0M_{\odot}$; Orosz et al. \cite{orosz}), and the more extreme merging black holes detected via gravitational waves, may form, if neutron star formation is strongly favoured for such comparatively massive progenitor stars.

\longtab{2}{
\begin{longtable}{lccccccll}
\caption{The stellar population of Westerlund 1}\\
\hline
\hline
ID         & RA (J2000)   & Dec (J2000)   & B  & V   & R   & I  &  Spectral Type   & Notes \\
\hline
\endfirsthead
\caption{continued.}\\
\hline
\hline
ID         & RA (J2000)   & Dec (J2000)   & B  & V   & R   & I  &  Spectral  Type  & Notes \\
\hline
\endhead
\hline
\endfoot
W1         &  16 46 59.28 & $-$45 50 46.7 & 21.9 & 18.37 & 16.09 & 13.65 & O9.5 Iab        & \\ 
W2a        &  16 46 59.71 & $-$45 50 51.1 & 20.4 & 16.69 & 14.23 & 11.73 & B2 Ia           & RV binary?$^2$, H$\alpha$ 
variable$^5$ \\
W4         &  16 47 01.42 & $-$45 50 37.1 & 18.7 & 14.47 & 11.80 & 9.15  & F3 Ia$^+$       & \\
W5         &  16 47 02.97 & $-$45 50 19.5 & 21.4 & 17.49 & 14.98 & 12.48 & WN10-11h/B0.5 Ia$^+$  & WR S, stripped primary$^{9}$ \\
W6a        &  16 47 03.04 & $-$45 50 23.6 & 22.2 & 18.41 & 15.80 & 13.16 & B0.5 Iab        & P(2.20d)$^{1}$, H$\alpha$ variable$^5$ \\   
W6b        &  16 47 02.93 & $-$45 50 22.3 & 23.6 & 20.20 & 17.91 & 15.25 & O9.5 III        & \\   
W7         &  16 47 03.62 & $-$45 50 14.2 & 20.0 & 15.57 & 12.73 & 9.99  & B5 Ia$^+$       & H$\alpha$ variable, Pulsator?$^5$ \\
W8a        &  16 47 04.79 & $-$45 50 24.9 & 19.9 & 15.50 & 12.64 & 9.89  & F8 Ia$^+$       & \\
W8b        &  16 47 04.95 & $-$45 50 26.7 & $-$  & $-$   & $-$   & $-$   & B1.5 Ia         & Pulsator?$^5$ \\
W9         &  16 47 04.14 & $-$45 50 31.1 & 21.8 & 17.47 & 14.47 & 11.74 & sgB[e]          & Interacting binary$^8$\\
W10        &  16 47 03.32 & $-$45 50 34.7 & $-$  & $-$   & $-$   & $-$   & B0.5 I+OB       & SB2 \\
W11        &  16 47 02.23 & $-$45 50 47.0 & 21.2 & 17.15 & 14.52 & 11.91 & B2 Ia           & \\
W12a       &  16 47 02.21 & $-$45 50 58.8 & 22.0 & 16.94 & 13.54 & 10.42 & F1 Ia$^+$       & \\
W13        &  16 47 06.45 & $-$45 50 26.0 & 21.1 & 17.19 & 14.63 & 12.06 & B0.5 Ia$^+$+OB  & E(9.27d)$^{1,3}$ \\
W14c       &  16 47 06.07 & $-$45 50 22.6 & $-$  & $-$   & $-$   & $-$   & WN5o            & WR R\\
W15        &  16 47 06.63 & $-$45 50 29.7 & 22.8 & 18.96 & 16.38 & 13.75 & O9 Ib           & \\
W16a       &  16 47 06.61 & $-$45 50 42.1 & 20.5 & 15.89 & 12.82 & 9.90  & A5 Ia$^+$       & H$\alpha$ variable$^5$ \\
W17        &  16 47 06.25 & $-$45 50 49.2 & 22.7 & 18.87 & 16.19 & 13.56 & O9 Iab          & \\
W18        &  16 47 05.71 & $-$45 50 50.5 & 21.2 & 17.32 & 14.81 & 12.27 & B0.5 Ia         & \\
W19        &  16 47 04.86 & $-$45 50 59.1 & 22.6 & 18.22 & 15.21 & 12.37 & B1 Ia           & H$\alpha$ variable$^5$\\
W20        &  16 47 03.09 & $-$45 52 18.8 & $-$  & $-$   & $-$   & $-$   & M5 Ia           & \\
W21        &  16 47 01.10 & $-$45 51 13.6 & 22.5 & 18.41 & 15.56 & 12.74 & B0.5 Ia         & Pulsator?$^5$ \\
W23a       &  16 47 02.57 & $-$45 51 08.7 & 22.1 & 17.85 & 14.91 & 12.07 & B2 Ia+B I?      & H$\alpha$ variable, Pulsator?$^5$ \\
W24        &  16 47 02.15 & $-$45 51 12.4 & 23.0 & 18.71 & 15.96 & 13.24 & O9 Iab          & Pulsator?$^5$ \\
W25        &  16 47 05.78 & $-$45 50 33.3 & 21.9 & 17.85 & 15.22 & 12.61 & O9 Iab          & \\
W26        &  16 47 05.40 & $-$45 50 36.5 & 22.1 & 16.79 & 12.63 & 9.19  & M2$\leftrightarrow$5 Ia & Spec. variable$^5$ \\
W27        &  16 47 05.15 & $-$45 50 41.3 & 21.5 & 17.94 & 15.35 & 12.80 & O7-8 Ia$^+$     & Merger remnant?$^{10}$\\
W28        &  16 47 04.66 & $-$45 50 38.4 & 20.9 & 16.87 & 14.26 & 11.64 & B2 Ia           & H$\alpha$ variable$^5$ \\
W29        &  16 47 04.41 & $-$45 50 39.8 & 22.6 & 18.66 & 16.02 & 13.38 & O9 Ib           & \\  
W30        &  16 47 04.11 & $-$45 50 39.0 & 22.4 & 18.45 & 15.80 & 13.20 & O4-5 Ia$^+$     & RV binary?$^{10}$ H$\alpha$ variable$^5$\\
W31        &  16 47 03.78 & $-$45 50 40.4 & $-$  & $-$   & $-$   & $-$   & B0 I+OB         & \\
W32        &  16 47 03.67 & $-$45 50 43.5 & $-$  & $-$   & $-$   & $-$   & F5 Ia$^+$       & \\
W33        &  16 47 04.12 & $-$45 50 48.3 & 20.0 & 15.61 & 12.78 & 10.04 & B5 Ia$^+$       & H$\alpha$ variable$^5$\\
W34        &  16 47 04.39 & $-$45 50 47.2 & 22.1 & 18.15 & 15.40 & 12.69 & B0 Ia           & \\
W35        &  16 47 04.20 & $-$45 50 53.5 & 22.7 & 18.59 & 16.00 & 13.31 & O9 Iab          & \\
W36        &  16 47 05.04 & $-$45 50 55.3 & 22.8 & 18.89 & 16.09 & 13.38 & OB Ia + OB Ia   & E(3.18d)$^1$, SB2$^7$, very broad Pa lines\\
W37        &  16 47 06.01 & $-$45 50 47.4 & 22.8 & 19.11 & 16.40 & 13.65 & O9 Ib           & \\
W38        &  16 47 02.86 & $-$45 50 46.0 & 23.2 & 19.10 & 16.47 & 13.81 & O9 Iab          & \\
W41        &  16 47 02.70 & $-$45 50 56.9 & 21.3 & 17.87 & 15.39 & 12.78 & O9 Iab          & \\
W42a       &  16 47 03.25 & $-$45 50 52.1 & $-$  & $-$   & $-$   & $-$   & B9 Ia$^+$       & H$\alpha$ variable, Pulsator?$^5$ \\
W43a       &  16 47 03.54 & $-$45 50 57.3 & 22.8 & 18.05 & 15.22 & 12.26 & B0 Ia           & RV(16.27d)$^4$, SB1, H$\alpha$ variable$^5$\\
W43b       &  16 47 03.52 & $-$45 50 56.5 & $-$  & $-$   & $-$   & $-$   & B1 Ia           & \\
W43c       &  16 47 03.76 & $-$45 50 58.3 & 20.4 & 18.35 & 16.18 & 13.66 & O9 Ib           & \\
W44        &  16 47 04.20 & $-$45 51 06.9 & 22.6 & 18.86 & 15.61 & 12.52 & WN9h:           & WR L, RV binary?$^5$ \\
W46a       &  16 47 03.91 & $-$45 51 19.5 & 23.0 & 18.55 & 15.46 & 12.46 & B1 Ia           & \\
W46b       &  16 47 03.61 & $-$45 51 20.0 & $-$  & $-$   & $-$   & $-$   & O9.5 Ib         & \\
W47        &  16 47 02.64 & $-$45 51 17.6 & 22.7 & 19.95 & 16.36 & 13.68 & O9.5 Iab        & \\
W49        &  16 47 01.90 & $-$45 50 31.5 & 22.6 & 18.76 & 16.30 & 13.80 & B0 Iab          & \\
W50b       &  16 47 01.17 & $-$45 50 26.7 & 22.8 & 19.66 & 17.21 & 14.69 & O9 III          & \\
W52        &  16 47 01.84 & $-$45 51 29.2 & 21.8 & 17.48 & 14.68 & 11.94 & B1.5 Ia         & P(6.7d)$^1$ \\
W53        &  16 47 00.48 & $-$45 51 32.0 & 22.9 & 18.51 & 15.80 & 13.13 & OB Ia + OB Ia   & P(1.30d)$^1$, SB2, very broad Pa lines\\
W54        &  16 47 03.06 & $-$45 51 30.5 & $-$  & $-$   & $-$   & $-$   & B0.5 Iab        & \\
W55        &  16 46 58.40 & $-$45 51 31.2 & 21.6 & 17.67 & 15.25 & 12.67 & B0 Ia           & \\
W56a       &  16 46 58.93 & $-$45 51 48.8 & 21.7 & 17.46 & 14.81 & 12.15 & B1.5 Ia         & \\
W56b       &  16 46 58.85 & $-$45 51 45.8 & 22.8 & 18.88 & 16.36 & 13.76 & O9.5 Ib         & \\
W57a       &  16 47 01.35 & $-$45 51 45.6 & 20.7 & 16.54 & 13.83 & 11.13 & B4 Ia           & Pulsator?$^5$ \\
W57c       &  16 47 01.59 & $-$45 51 45.5 & $-$  & $-$   & $-$   & $-$   & WN7o            & WR P \\
W60        &  16 47 04.13 & $-$45 51 52.1 & 22.8 & 18.50 & 15.96 & 13.28 & B0 Iab          & \\
W61a       &  16 47 02.29 & $-$45 51 41.6 & 21.2 & 17.16 & 14.62 & 12.01 & B0.5 Ia         & H$\alpha$ variable$^5$ \\
W61b       &  16 47 02.56 & $-$45 51 41.6 & 22.7 & 18.59 & 16.00 & 13.31 & O9.5 Iab        & \\
W62a       &  16 47 02.51 & $-$45 51 37.9 & $-$  & $-$   & $-$   & $-$   & B0.5 Ib         & \\
W63a       &  16 47 03.39 & $-$45 51 57.7 & 22.6 & 18.56 & 16.20 & 13.68 & B0 Iab          & \\
W65        &  16 47 03.89 & $-$45 51 46.3 & 22.9 & 18.73 & 16.27 & 13.68 & O9 Ib           & \\
W66        &  16 47 03.96 & $-$45 51 37.5 & $-$  & 19.79 & 16.85 & 13.96 & WC9d            & WR M \\
W70        &  16 47 09.36 & $-$45 50 49.6 & 21.2 & 16.88 & 14.10 & 11.29 & B3 Ia           & H$\alpha$ variable$^5$\\
W71        &  16 47 08.44 & $-$45 50 49.3 & 21.5 & 17.01 & 14.06 & 11.16 & B2.5 Ia         & H$\alpha$ variable, Pulsator?$^5$ \\
W72        &  16 47 08.32 & $-$45 50 45.5 & $-$  & 19.69 & 16.59 & 13.68 & WN7b            & WR A, P(7.63d)$^1$ \\
W74        &  16 47 07.08 & $-$45 50 13.1 & $-$  & $-$   & $-$   & $-$   & O9.5 Iab        & \\
W75        &  16 47 08.93 & $-$45 49 58.4 & $-$  & $-$   & $-$   & $-$   & M4 Ia           & \\
W78        &  16 47 01.54 & $-$45 49 57.8 & 21.0 & 17.06 & 14.54 & 12.04 & B1 Ia           & Pulsator?$^5$ \\
W84        &  16 46 59.03 & $-$45 50 28.2 & 21.3 & 17.82 & 15.60 & 13.63 & O9.5 Ib         & \\
W86        &  16 46 57.15 & $-$45 50 09.9 & 22.9 & 18.76 & 16.43 & 14.00 & O9.5 Ib         & \\
W228b      &  16 46 58.05 & $-$45 53 01.0 & $-$  & $-$   & $-$   & $-$   & O9 Ib           & \\
W232       &  16 47 01.41 & $-$45 52 34.9 & 21.3 & 17.53 & 15.25 & 12.85 & B0 Iab          & \\
W237       &  16 47 03.09 & $-$45 52 18.8 & 22.8 & 17.49 & 13.00 & 9.19  & M3 Ia           & Spec. variable$^5$ \\
W238       &  16 47 04.41 & $-$45 52 27.6 & 21.4 & 17.47 & 14.98 & 12.45 & B1 Iab          & \\
W239       &  16 47 05.21 & $-$45 52 25.0 & 21.7 & 17.86 & 15.39 & 12.90 & WC9d            & WR F, RV(5.05d)$^6$, SB1 \\ 
W241       &  16 47 06.06 & $-$45 52 08.3 & $-$  & $-$   & $-$   & $-$   & WC9             & WR E, RV binary?$^2$ \\
W243       &  16 47 07.55 & $-$45 52 28.5 & $-$  & $-$   & $-$   & $-$   & LBV             & Spec. variable, pulsator$^2$\\
W265       &  16 47 06.26 & $-$45 49 23.7 & 22.0 & 17.05 & 13.62 & 10.54 & F1$\leftrightarrow$5 Ia$^+$     & Spec. variable$^5$, pulsator$^3$ \\
W373       &  16 46 57.71 & $-$45 53 20.1 & $-$  & $-$   & $-$   & $-$   & B0 Iab          & \\  
           &              &               &      &       &       &       &                 & \\
WR B       &  16 47 05.36 & $-$45 51 05.0 & $-$  & 20.99 & 17.50 & 14.37 & WN7o            & E(3.51d)$^1$, SB1 \\
WR C       &  16 47 04.40 & $-$45 51 03.8 & $-$  & $-$   & $-$   & $-$   & WC9d            & \\
WR D       &  16 47 06.24 & $-$45 51 26.5 & $-$  & $-$   & $-$   & $-$   & WN7o            & \\      
WR G       &  16 47 04.01 & $-$45 51 25.2 & 22.7 & 20.87 & 17.75 & 14.68 & WN7o            & \\
WR H       &  16 47 04.22 & $-$45 51 20.2 & $-$  & $-$   & $-$   & $-$   & WC9d            & \\
WR I       &  16 47 00.88 & $-$45 51 20.8 & $-$  & $-$   & $-$   & $-$   & WN8o            & \\
WR J       &  16 47 02.47 & $-$45 51 00.1 & $-$  & $-$   & $-$   & $-$   & WN5h            & \\
WR K       &  16 47 03.25 & $-$45 50 43.8 & $-$  & $-$   & $-$   & $-$   & WC8             & \\  
WR N       &  16 46 59.9  & $-$45 55 26   & $-$  & $-$   & 16.90 & 13.00 & WC9d            & \\
WR O       &  16 47 07.66 & $-$45 52 35.9 & $-$  & $-$   & $-$   & $-$   & WN6o            & \\
WR Q       &  16 46 55.55 & $-$45 51 35.0 & $-$  & 20.30 & 17.50 & 14.70 & WN6o            & \\
WR T       &  16 46 46.3  & $-$45 47 58   & $-$  & $-$   & $-$   & $-$   & WC9d            & \\
WR U       &  16 47 06.55 & $-$45 50 39.0 & $-$  & $-$   & $-$   & $-$   & WN6o            & \\
WR V       &  16 47 03.81 & $-$45 50 38.8 & $-$  & $-$   & $-$   & $-$   & WN8o            & \\
WR W       &  16 47 07.58 & $-$45 49 22.2 & $-$  & $-$   & $-$   & $-$   & WN6h            & \\ 
WR X       &  16 47 14.1  & $-$45 48 32   & $-$  & $-$   & $-$   & $-$   & WN5o            & \\
           &              &               &      &       &       &       &                 & \\
1001       & 16 46 49.20  & $-$45 53 10.0 & $-$  & $-$   & $-$   & $-$   & O+O?            & very Pa broad lines \\
1002       & 16 46 49.68  & $-$45 52 53.0 & $-$  & {\em 18.89}   & {\em 16.59} & {\em 14.41}& O9-9.5 II+O? & P(0.144d)$^1$, broad Pa 
lines \\
1003       & 16 46 52.32  & $-$45 52 03.4 & $-$  & $-$   & $-$   & $-$   & O9-9.5 bin?     & broad Pa lines \\
1004       & 16 46 53.52  & $-$45 53 00.2 & $-$  & {\em 18.06}   & {\em 16.60} & {\em 14.43}& OeBe star      & \\
1005       & 16 46 54.24  & $-$45 51 54.7 & 23.6 & 18.93 & 16.08 & 13.26 & B0 Iab          & W3002 \\
1006       & 16 46 54.48  & $-$45 53 30.1 & $-$  & $-$   & {\em 17.41} & {\em 14.98} & O9-9.5 III bin? & P(0.127d)$^1$, broad Pa lines\\
1007       & 16 46 54.96  & $-$45 50 06.0 & $-$  & $-$   & $-$   & $-$   & O9-9.5 III      & \\
1008       & 16 46 55.44  & $-$45 51 54.4 & 23.1 & 19.73 & 17.11 & 14.40 & O9.5 II         & \\
1009       & 16 46 55.92  & $-$45 51 41.4 & 22.9 & 19.07 & 16.74 & 14.08 & B0 Ib           & W2002 \\
1010       & 16 46 55.92  & $-$45 52 10.2 & $-$  & $-$   & $-$   & $-$   & O+O?            & very broad Pa lines\\
1011       & 16 46 56.86  & $-$45 52 04.4 & 22.5 & 19.14 & 17.10 & 14.79 & O+O?            & very broad Pa lines\\
1012       & 16 46 56.95  & $-$45 50 56.0 & 23.8 & 20.36 & 17.78 & 14.97 & O9-9.5 III bin? & broad Pa lines \\
1013       & 16 46 57.60  & $-$45 52 30.7 & $-$  & $-$   & $-$   & $-$   & O+O?            & very Pa broad lines\\
1014       & 16 46 57.81  & $-$45 51 19.8 & 23.3 & 19.58 & 17.40 & 15.03 & O9-9.5 III bin? & broad Pa, H$\alpha$/He~I double? \\
1015       & 16 46 57.96  & $-$45 51 40.7 & 23.1 & 19.24 & 16.77 & 14.17 & O9 III          & \\
1016       & 16 46 58.08  & $-$45 52 46.9 & $-$  & $-$   & $-$   & $-$   & O9-9.5 III bin? & broad Pa lines\\
1017       & 16 46 58.23  & $-$45 50 33.9 & 23.5 & 19.91 & 17.44 & 14.94 & O9-9.5 III bin? & broad Pa lines\\
1018       & 16 46 58.32  & $-$45 50 56.8 & $-$  & $-$   & $-$   & $-$   & O9.5 Iab & CXO164658.2-455056 \\
1019       & 16 46 58.36  & $-$45 51 48.8 & 23.8 & 20.33 & 17.86 & 15.32 & O9-9.5 III bin? & broad Pa lines\\
1020       & 16 46 58.48  & $-$45 52 27.1 & 21.8 & 18.45 & 16.38 & 14.19 & O9-9.5+O?       & broad Pa lines \\
1021       & 16 46 58.78  & $-$45 54 31.9 & $-$  & {\em 20.31} & {\em 17.38} & {\em 14.82}   & O9-9.5III bin & P(4.43d), SB2, broad Pa lines \\
1022       & 16 46 59.93  & $-$45 50 25.4 & 23.0 & 19.48 & 17.08 & 14.82 & O9.5 II         & P(0.1703d)$^1$\\
1023       & 16 47 00.24  & $-$45 51 10.4 & $-$  & $-$   & $-$   & $-$   & O9 III          & \\
1024       & 16 47 00.72  & $-$45 51 01.8 & 23.3 & 19.38 & 16.72 & 14.04 & O9.5 Iab      & W2011 \\
1025       & 16 47 00.76  & $-$45 52 04.6 & 22.9 & 19.91 & 17.65 & 15.26 & O+O?            & very broad Pa lines\\
1026       & 16 47 00.96  & $-$45 49 48.7 & $-$  & $-$   & $-$   & $-$   & O9-9.5 III      & \\
1027       & 16 47 00.96  & $-$45 50 06.7 & $-$  & $-$   & $-$   & $-$   & O9.5 Iab & CXO164701.0-455006 \\
1028       & 16 47 01.32  & $-$45 51 38.2 & 23.9 & 20.76 & 17.98 & 15.29 & O9-9.5 III bin? & broad Pa lines, H$\alpha$/He~I double?\\
1029       & 16 47 01.44  & $-$45 49 50.2 & $-$  & $-$   & $-$   & $-$   & O9-9.5 III bin? & broad Pa lines\\
1030       & 16 47 01.69  & $-$45 52 57.8 & $-$  & $-$   & {\em 15.39} & {\em 12.97} & O9.5 Iab & W3005 \\
1031       & 16 47 01.92  & $-$45 50 56.4 & $-$  & $-$   & $-$   & $-$   & O9 III & \\
1032       & 16 47 02.27  & $-$45 50 17.6 & 24.0 & 21.18 & 18.01 & 15.38 & O9-9.5 III bin? & broad Pa lines, H$\alpha$/He~I double?\\
1033       & 16 47 02.40  & $-$45 52 34.3 & 22.6 & 18.96 & 16.43 & 13.81 & O9-9.5 I-III    & C07-X5, skewed lines?\\
1034       & 16 47 02.52  & $-$45 51 48.2 & 22.5 & 18.69 & 16.34 & 13.85 & O9.5 Iab        & \\
1035       & 16 47 02.64  & $-$45 51 51.1 & 23.5 & 19.56 & 17.15 & 14.54 & O9-9.5 III bin? & broad Pa lines\\
1036       & 16 47 02.77  & $-$45 52 12.4 & 23.5 & 19.29 & 16.69 & 14.09 & O9.5 Iab         & C07-X4, W2017 \\
1037       & 16 47 02.84  & $-$45 50 06.3 & 24.0 & 19.62 & 17.00 & 14.41 & O9.5 II         & \\
1038       & 16 47 03.60  & $-$45 48 57.2 & $-$  & $-$   & $-$   & $-$   & O9 III & \\
1039       & 16 47 03.60  & $-$45 49 21.7 & $-$  & $-$   & $-$   & $-$   & B1 Ia  & W3019 \\
1040       & 16 47 04.52  & $-$45 50 08.8 & 21.9 & 18.89 & 16.37 & 13.76 & O9-9.5 I-III bin? & W2019, C07-X3, broad Pa 
lines \\
1041       & 16 47 04.56  & $-$45 51 09.4 & $-$  & $-$   & $-$   & $-$   & O9.5 Iab bin? & CXO164704.4-455109, broad Pa lines \\
1042       & 16 47 04.56  & $-$45 52 06.6 & 22.8 & 19.17 & 16.80 & 14.27 & O9.5 II         &  \\
1043       & 16 47 04.57  & $-$45 50 59.3 & 23.3 & 20.17 & 17.39 & 14.53 & O9.5 II-III     &  \\
1044       & 16 47 05.56  & $-$45 49 51.5 & 24.1 & 20.28 & 17.85 & 15.36 & O9-9.5 III bin? & broad Pa lines, H$\alpha$ infilled?\\
1045       & 16 47 05.82  & $-$45 51 54.9 & 23.2 & 19.77 & 17.20 & 14.49 & O9.5 II         & \\
1046       & 16 47 06.00  & $-$45 49 56.9 & 22.5 & 18.87 & 16.59 & 14.19 & O+O?            & very Pa broad lines, He~I double?\\
1047       & 16 47 06.11  & $-$45 52 32.1 & 22.1 & 18.93 & 16.58 & 14.03 & O9.5 II         & \\
1048       & 16 47 06.26  & $-$45 51 03.9 & 23.8 & 20.50 & 16.47 & 13.80 & B1.5 Ia         & P(5.2d)$^1$ \\ 
1049       & 16 47 06.65  & $-$45 47 38.5 & 22.1 & 17.70 & 14.81 & 12.05 & B1-2 Ia$^+$     & \\ 
1050       & 16 47 06.77  & $-$45 49 55.3 & 23.1 & 19.52 & 16.97 & 14.44 & O9.5 II         & \\
1051       & 16 47 06.96  & $-$45 49 40.1 & $-$  & $-$   & {\em 16.70} & {\em 13.94} & O9 III & \\
1052       & 16 47 06.96  & $-$45 52 55.9 & $-$  & $-$   & $-$   & $-$   & O9 III & \\
1053       & 16 47 07.44  & $-$45 48 50.0 & $-$  & $-$   & $-$   & $-$   & B0 Ib  & \\
1054       & 16 47 07.68  & $-$45 51 41.0 & $-$  & $-$   & $-$   & $-$   & O9-9.5 bin?     & broad Pa lines\\
1055       & 16 47 07.92  & $-$45 51 47.9 & $-$  & $-$   & {\em 16.94} & {\em 13.94} & B0 Ib (+O?) & W3024, broad Pa lines \\
1056       & 16 47 08.64  & $-$45 51 01.4 & 24.2 & 20.54 & 17.31 & 14.30 & O9.5 II         & \\
1057       & 16 47 08.67  & $-$45 50 47.1 & 23.3 & 19.39 & 16.60 & 13.71 & O9.5-B0 Iab      & W2028, H$\alpha$ infilled, C~II em.\\
1058       & 16 47 08.88  & $-$45 51 24.5 & $-$  & $-$   & $-$   & $-$   & O9 III & \\
1059       & 16 47 09.12  & $-$45 53 20.8 & $-$  & $-$   & $-$   & $-$   & O9 III?& \\
1060       & 16 47 09.18  & $-$45 50 48.3 & 23.3 & 20.09 & 17.09 & 14.31 & O9.5 II         & \\
1061       & 16 47 09.74  & $-$45 50 40.2 & 23.9 & 20.91 & 18.16 & 15.42 & O9-9.5 III bin? & broad Pa lines\\
1062       & 16 47 10.62  & $-$45 50 46.4 & 23.6 & 20.03 & 17.48 & 14.90 & O+O?            & very broad Pa lines\\
1063       & 16 47 10.80  & $-$45 49 47.6 & $-$  & $-$   & $-$   & $-$   & O9 III & \\
1064       & 16 47 11.52  & $-$45 49 59.5 & $-$  & $-$   & $-$   & $-$   & O9.5 Iab      & CXO164711.5-455000 \\
1065       & 16 47 11.60  & $-$45 49 22.4 & 23.1 & 19.08 & 16.19 & 13.28 & B0 Ib  & W3003, RV(11.12d)$^4$, SB1 \\
1066       & 16 47 12.72  & $-$45 50 55.3 & $-$  & $-$   & $-$   & $-$   & O9 III & \\
1067       & 16 47 13.39  & $-$45 49 10.5 & 23.2 & 18.93 & 15.88 & 12.98 & B0 Iab & W3004 \\
1068       & 16 47 16.56  & $-$45 51 41.0 & $-$  & $-$   & $-$   & $-$   & B0Ib & blended \\
1069       & 16 47 24.24  & $-$45 53 29.0 & $-$  & $-$   & $-$   & $-$   & B5 Ia$^+$ & \\   
\end{longtable}
{Compilation of stellar classifications for members of Westerlund 1. Column 1 lists the primary optical identifier 
for the sources and columns 2 and 3 their co-ordinates. Columns 4-7 present $B-$, $V-$, $R-$ and $I-$band photometry 
derived from the dataset described in Clark et al. (\cite{clark05}) and, if not available, Bonanos (\cite{bonanos}; italics).
Column 8 presents the spectral classification and where appropriate column nine presents other designations (including  the  $W2xxx$ and $W3xxx$ designations  previously employed but superceded by this work) and notes regarding spectral appearance and 
variability, with {\bf E}clipsing and  {\bf P}eriodic photometric variables and {\bf R}adial {\bf V}elocity spectroscopic variables listed with their relevant periods. The vast majority of classifications derive from this work, Clark et al. (\cite{clark05}), 
Crowther et al. (\cite{crowther06}) and Negueruela et al. (\cite{neg10}), with exceptions highlighted.
Additional data used in the construction of this table from $^1$Bonanos (\cite{bonanos}), $^2$Ritchie et al. (\cite{ritchie09a}), 
$^3$Ritchie et al (\cite{ritchie10}), $^4$Ritchie et al. (\cite{ritchie11}), $^5$Clark et al. (\cite{clark10}), 
$^6$Clark et al. (\cite{clark11}) $^7$Koumpia \& Bonanos (\cite{kb12}), $^8$Clark et al. (\cite{clark13}) $^{9}$Clark et al. 
(\cite{clark14b}) and $^{10}$Clark et al. (\cite{clark19a}).  
}}

\begin{table}
\caption{Field stars}
\begin{center}
\begin{tabular}{llll}
\hline
\hline
ID & Spectral Type           &  RA (J2000) & Dec (J2000) \\
\hline
F2 & M0--2 bin & 16 46 51.36 & -45 50 10.3 \\
F3 & B~V       & 16 46 54.00 & -45 53 06.4 \\
F4 & M0--2     & 16 46 54.24 & -45 49 42.6 \\
F5 & G5--K5    & 16 46 55.92 & -45 52 18.8 \\
F6 & G5--K5    & 16 46 57.12 & -45 51 36.0 \\
F1 & M2~II-III & 16 46 57.84 & -45 52 18.5 \\
F7 & K0--K5    & 16 47 07.44 & -45 48 42.8 \\
F8 & K0--K5    & 16 47 18.00 & -45 49 43.7 \\
\hline
\end{tabular}
\end{center}
\end{table}

\begin{table}
\caption{Summary of the WR populations of  Wd1 and the Arches, Quintuplet and 
Galactic Centre clusters}
\begin{center}
\begin{tabular}{lcccc}
\hline
\hline
          &  Wd1  & Arches & Quint. & Gal Cen. \\
Sub-type  &       &        &        &          \\
\hline
WN8-9ha   &   0   & 13  &2 & 0 \\
          &       &     &  &   \\
WN5       &   3   & 0 & 0 & 0 \\ 
WN6       &   4   & 0 &1 & 1 \\
WN7       &   5   & 0 &0 & 3 \\
WN8       &   2   & 0 & 0 & 5 \\
WN9-11h   &   2   & 0 &3 & 8 \\
          &       &   & &   \\
WN/WC     &   0   & 0 & 0 & 2 \\
          &       &   & &   \\
WC5/6     &   0   & 0 & 0 & 1 \\
WC8-9     &   8   & 0 & 14& 13\\
\hline
\end{tabular}
\end{center}
{WR populations for the Arches, Quintuplet and Galactic Centre clusters derive 
from Clark et al. (\cite{clark18a}, \cite{clark18b}), Paumard et al. (\cite{paumard}) and
Martins et al. (\cite{martins07}). For ease of presentation the Galactic 
Centre WN5/6 star IRS 16SE2 is listed as WN6 in this table. Finally, we also include the 
WC9d star CXOGC J174617.7-285007 (Mauerhan et al. \cite{mauerhan}), which is located to 
the south of the Quintuplet in the total for this sub-type  for this cluster.} 
\end{table}
 
\begin{acknowledgements}
We are indebted to Prof. Simon Goodwin for calculating the integrated cluster masses discussed
in Sect.  4.2.
This research is partially supported by the Spanish Government under grants
AYA2015-68012-C2-2-P and PGC2018-093741-B-C21 (MICIU/AEI/FEDER, UE), and 
made use of the SIMBAD database, operated
at CDS, Strasbourg, France.

\end{acknowledgements}

\end{document}